# Mapping the Landscape of AI-Driven Human Resource Management: A Social Network Analysis of Research Collaboration

**Mehrdad Maghsoudi[1], Motahareh Kamrani Shahri[2], Mehrdad Agha Mohammad Ali Kermani[2], and Rahim Khanizad[2]**

[1] Department of Industrial and Information Management, Faculty of Management and Accounting, Shahid Beheshti University, Tehran, Iran
[2] Department of Management, Economics and Progress Engineering, Iran University of Science and Technology, Iran, Tehran, Iran

Corresponding author: Mehrdad Agha Mohammad Ali Kermani (e-mail: M_Kermani@iust.ac.ir ).

**ABSTRACT** As artificial intelligence (AI) transforms human resource management (HRM), understanding the research landscape becomes crucial for both academics and practitioners. While existing studies examine isolated aspects of AI in HRM, a comprehensive analysis of collaboration patterns and emerging themes remains lacking. This research employs social network analysis (SNA) to examine the co-authorship network within AI applications in HRM research, providing insights into collaboration dynamics and identifying key research directions. Through analysis of centrality measures and application of the TOPSIS method, the study identifies influential authors, institutions, and emerging research themes. Analysis of 102,296 authors and 287,799 collaborations reveals distinct communities focusing on specific aspects of AI-HRM across regions. The findings identify four primary research themes: AI for System Identification and Control, focusing on workforce planning and adaptive management; HR Analytics and Performance Management, emphasizing data-driven decision making; Machine Learning for Classification and Prediction, addressing talent acquisition and retention; and AI-Driven HR Decision-Making, exploring strategic planning and unbiased evaluation systems. The country co-authorship network analysis uncovers three main communities: Global HR Applications, HRM in the Middle East and Asia, and Global Integration of AI in HRM, reflecting shared regional challenges. Institutional collaboration patterns indicate five distinct communities, from established Asian AI research centers to emerging research hubs in developing economies. These findings provide valuable insights for researchers exploring collaboration opportunities, practitioners implementing AI solutions, and policymakers developing strategic frameworks for AI adoption in HRM. This research contributes to understanding the evolving landscape of AI-HRM research and offers practical guidelines for leveraging AI in HR practices.

## I. INTRODUCTION

The human resources (HR) function is undergoing a profound transformation, driven by emerging economic, political, social, and technological trends. Increasingly, HR is shifting from a primarily administrative role to that of a strategic business partner, essential for achieving organizational excellence and maintaining competitive advantage [1, 2]. Strategic human resource management (SHRM) involves aligning HR practices and policies with organizational goals to enhance long-term success. It entails HR's role in business strategy by concentrating on enhancing human capital, organizational culture, leadership, and employee engagement to drive performance and innovation[3]. While traditional HR practices have focused on recruitment, selection, performance management, and retention, there is now a broader emphasis on strategic responsibilities, including fostering interpersonal interactions within organizations [4].

Artificial Intelligence (AI) has become a pivotal technology in this transformation, automating routine administrative tasks and significantly enhancing HR operations. By facilitating knowledge sharing and collaboration, AI is improving employee experiences and increasing HR effectiveness [5, 6]. However, to fully leverage AI's transformative potential, it is crucial to identify and develop the capabilities and competencies required for managers to successfully implement and oversee AI-driven HR initiatives [7].

Despite growing interest in AI within HR research, a substantial gap remains in the literature regarding comprehensive analyses of AI's impact on Human Resource Management(HRM) [8, 9]. Although AI's potential to revolutionize HRM is well-recognized, particularly in decision-making through data-driven algorithms , there is a pressing need for deeper insight into how AI can be integrated into HR sub-functions [10, 11]. Additionally, research has yet to fully explore how AI-enabled HRM functions can be aligned with other operational tasks to enhance overall employee outcomes [7, 12].

Recent studies have begun to examine AI applications in HRM, yet there remains a scarcity of research that provides a thorough overview of trends in this field [13, 14]. This paper aims to address this gap by offering a comprehensive analysis of AI's role in HRM through a novel methodology—social network analysis (SNA). Unlike previous studies that relied on bibliometric analysis, this research employs SNA to uncover the complex relationships within AI applications in HRM. By analyzing co-authorship networks, this study reveals overlooked connections among authors, institutions, keywords, and countries, identifying emerging topics and potential knowledge hubs previously missed in the literature.

Social Network Analysis is particularly valuable as it allows for an in-depth examination of the interactions and collaborations within social networks. By identifying key figures and influential entities, SNA helps to understand how network connections influence behavior, beliefs, and performance among groups and individuals [15, 16]. In the context of AI in HRM, SNA provides insights into the dynamics of author collaborations, highlighting influential contributors and their roles in advancing research. Furthermore, by mapping contributions from notable institutions and countries, this study fosters global cooperation in AI-driven HRM research.

This study provides a unique contribution to the AI-HRM research landscape by combining Social Network Analysis (SNA) and TOPSIS to map research collaboration patterns and emerging themes. Unlike previous studies, which have often focused on isolated aspects of AI in HRM, this research offers a comprehensive approach that identifies key authors, institutions, and research hubs globally. The novelty lies in the integration of network analysis with decision-making models, uncovering previously overlooked patterns in collaboration. The study not only advances theoretical knowledge but also offers actionable insights for practitioners, policymakers, and academics involved in AI-driven HR practices.

This research seeks to answer the following key questions to enhance the understanding of AI's impact on HRM:

1. Who are the most influential authors in the HRM and AI co-authoring network?
2. Is the collaboration between countries in co-authorship driven by shared challenges, and what underpins scientific collaboration among universities and institutions in HRM and AI?
3. What underpins scientific collaboration among universities and institutions in HRM and AI?
4. Does the analysis of keyword networks in HRM and AI research reveal distinct and coherent patterns?

The structure of this paper is organized as follows. Chapter 2 reviews the literature. Chapter 3 outlines the research methodology. Chapter 4 presents the research findings. Chapter 5 discusses the results, providing explicit answers to the research questions, managerial implications, and limitations, and future research directions. Finally, Chapter 6 concludes the study.

## II. LITERATURE REVIEW

This section reviews the literature on AI implementation within HRM and the use of SNA in researching collaboration. The review centers on the impact of AI on HR practices and the application of SNA to understand the nature of co-authorship networks. Synthesizing existing knowledge in this regard, this section intends to provide a good understanding of the role AI plays in modern HR practices and the nature of research collaboration in this field.

### *Application of AI in HRM*

The integration of AI into HRM necessitates a comprehensive understanding of its capabilities and applications within the HR function [17]. AI, a field simulating human intelligence in machines, encompasses a range of technologies including machine learning, natural language processing, and predictive analytics [9]. When applied to HRM, AI automates processes, informs decision-making, and enhances employee experiences [18]. By analyzing vast datasets, AI algorithms generate insights, predictions, and recommendations to improve efficiency, reduce bias, and personalize HR interactions [19]. This integration marks a transformative shift, enabling data-driven decision-making, increased productivity, and improved employee satisfaction [17]. While AI offers significant advantages, challenges such as ethical considerations, data privacy, and the need for ongoing human expertise must be carefully addressed to ensure its responsible implementation and maximize its benefits [20].

AI is fundamentally transforming HRM practices [21]. AI-driven tools are reshaping talent acquisition by streamlining processes and enhancing candidate selection [22]. Automated systems are employed to efficiently filter and assess large applicant pools, identifying candidates aligned with organizational needs [23]. Moreover, predictive analytics, leveraging historical data, forecasting candidate performance, and optimizing the hiring process [24]. In the realm of employee learning and development, AI-powered platforms offer personalized training experiences, adapting content to individual learning styles and knowledge levels [25].

AI also revolutionizes performance management by enabling continuous feedback, predictive analytics, and data-driven performance evaluations [23]. Additionally, AI-powered chatbots and sentiment analysis tools facilitate employee engagement by fostering instant communication and monitoring workplace morale [18]. Predictive analytics, driven by AI, offer valuable insights for talent retention, workforce planning, and succession management, enabling HR to proactively address organizational needs [26]. These AI applications transform HRM by fostering agility, strategic foresight, and a strong focus on employee well-being, thereby positioning organizations to thrive in the dynamic workforce landscape [27].

Strategic Human Resource Management (SHRM) traditionally encompasses six core functions: business administration, workforce planning, human capital development, compensation, labor relations, and risk management [28]. These domains are essential for achieving organizational objectives [29]. AI has the potential to enhance these HRM functions significantly [24]. For instance, in recruitment, AI supports HR managers by automating tasks such as filtering and preselection, thereby identifying candidates who best meet job requirements [29, 30]. Figure 1 illustrates the types and extent of AI applications in key HRM areas: Acquisition, Development, and Retention [31].

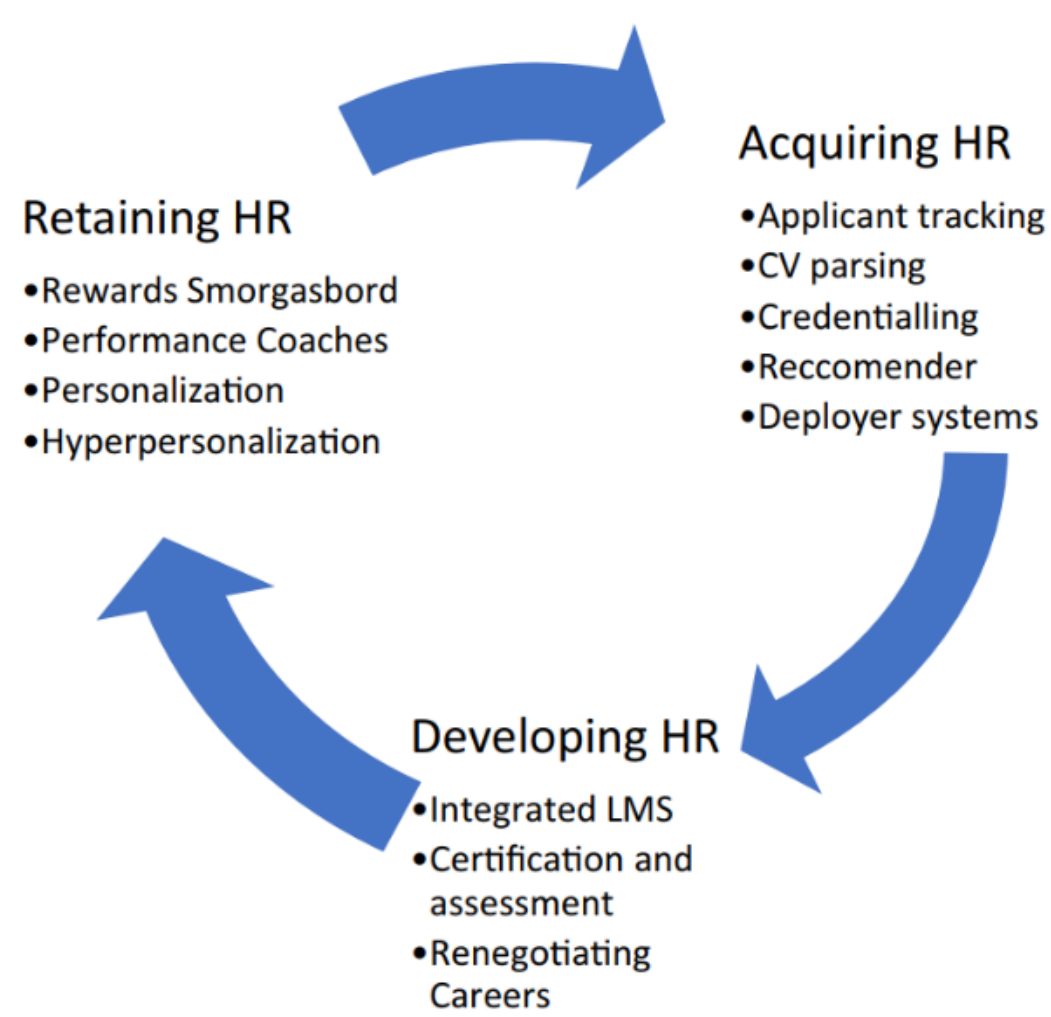

**Figure 1 AI and strategic HRM practice choice menus [32]**

AI can also significantly improve HR efficiency and quality through the use of chatbots, a common AI application in HR today [33, 34]. Table 1 outlines the potential impacts of AI on HRM in public sector organizations, both currently and in the future [35].

**Table 1 HRM and AI**

| HR Tasks | Potential Benefits of AI-Enabled Tools |
|---|---|
| Talent Acquisition | Increased processing capacity and more precise candidate evaluation by expanding the applicant pool |
| Employee Development | Immediate delivery of personalized training and onboarding |
| Performance Management | Comprehensive employee performance assessment, compared against peers and various metrics |
| Compensation | Automatic tracking of compensation across multiple data sources |
| Turnover and Retention | Strategic insights into incentives and potential employee departures |

AI is emerging as a transformative force within HRM, with its applications extending across various HR functions [21]. For example, AI-powered tools are enhancing talent acquisition by automating candidate identification and evaluation processes [36]. Advanced algorithms analyze resumes and social media profiles to match candidates with appropriate roles, while applicant tracking systems (ATS) streamline administrative tasks, boosting efficiency [37, 38]. Job aggregators and career platforms use AI to consolidate job openings and identify potential talent pools [27, 39].

Beyond recruitment, AI is reshaping employee development by supplementing traditional training methods with AI-driven learning platforms [40]. These platforms offer personalized, data-driven learning experiences by analyzing vast datasets to identify training needs, measure learning outcomes, and optimize training programs [18, 27].

Analyzing co-authorship networks through Social Network Analysis (SNA) is vital for understanding academic collaborations, trends, and influential experts. SNA quantifies collaborative intensity, identifies central figures, and visualizes connectivity within research networks. Introduced in the mid-1900s and popularized by "small world" experiments, SNA is crucial for comprehending complex social phenomena, assessing network characteristics, and mapping collaboration landscapes in various industries, including research[41].

### *Related works*

The Related Works section focuses specifically on reviewing papers that have conducted comprehensive and systematic analyses of AI applications in HRM research. Rather than examining all individual papers exploring various aspects of AI in HRM, this section analyzes those studies that, similar to the present research, have undertaken collective reviews of AI-HRM literature. This approach provides a meta-level understanding of how researchers have systematically analyzed and synthesized knowledge in this field. The selected papers represent significant contributions in systematically reviewing and analyzing multiple studies, rather than individual implementations or specific applications of AI in HRM.

Research on AI applications in HRM has rapidly increased in recent years, enhancing HR functions and organizational outcomes. [22] identified a significant lack of in-depth, organized research on HRM (AI) in current studies, particularly in the development of a multilevel framework that could serve as a foundation for future research by scholars. To address this disparity, they suggest a comprehensive framework that provides a foundation for upcoming researchers to establish connections between various variables, beginning at the contextual level and leading to HRM and organizational outcomes that ultimately improve the operational and financial performance of organizations.

AI and AI-based applications are increasingly used in HRM in both local and global organizations, leading to new research on topics like the social presence of AI, its effects on outcomes, and evaluating practices. Further research is needed for a comprehensive understanding and future directions. [42] present a systematic review of the theme of this special issue and offer a nuanced understating of what is known, yet to be known, and future research directions to frame a future research agenda for international HRM.

Research on AI in HRM has gained attention, but few studies address the trends and competencies needed for its adoption. [25] have a systematic review and bibliometric analysis identified managerial capabilities required for AI adoption in HRM, using the Dynamic Capabilities View. Recruitment and selection are common areas for AI use. Managerial cognitive, human, and social capital are important for AI adoption. Multiple methods were used to explore AI adoption in HRM.

Next research examines how AI has revolutionized HR functions, from talent acquisition to performance management. The study utilizes a bibliometric approach to analyze research trends between 2013 and 2022, focusing on identifying key research clusters and geographical patterns. By employing VOS viewer, the authors mapped out the intellectual structure of the field, revealing India as a leading contributor and highlighting areas such as machine learning for resource management, AI for recruitment, and AI-driven training as emerging research frontiers[43].

HRM is changing its focus from traditional methods like recruitment, training, and development to more advanced techniques like automation, AI integration, and augmented intelligence. [44] explore the application of AI in different activities associated with HRM. The research identifies the particular AI technologies utilized in HRM, investigates the advantages and difficulties of incorporating AI in HR, and the strategies applied to address these hurdles. The incorporation of AI in HRM is a new trend and only a small number of businesses have successfully adopted it in every HR process.

By systematically reviewing existing literature, the research explores how AI and automation are reshaping Human Resource Development (HRD) practices. The study identified key HRD processes impacted by these technologies and analyzed their influence on overall HRD outcomes. Findings offer insights into the challenges and opportunities presented by AI and automation, contributing to the ongoing discourse on the future of HRD [45].

This research employs Gephi, a specialized social network analysis tool, for analyzing co-authorship networks. While various software packages exist for network analysis, Gephi offers distinctive analytical capabilities crucial for this study. The software provides robust computation of multiple centrality metrics (degree, betweenness, closeness, and eigenvector centrality) to identify key actors in complex networks. Its advanced community detection algorithms, combined with unique filtering features, enable detailed analysis of specific communities and their network metrics - a capability particularly valuable for understanding research group dynamics.

These specialized features facilitate comprehensive examination of scholarly collaboration in the AI-HRM domain. Through co-authorship pattern analysis, the research identifies key contributors within specific research communities, reveals knowledge flow patterns, and maps emerging research clusters. The ability to filter and analyze specific communities while computing their distinct network metrics provides deeper insights into collaboration patterns. By examining author relationships, institutional connections, and collaboration dynamics, this analysis offers a thorough understanding of the field's structure and evolution. Table 2 provides a summary of the topic, method, and results of past research.

**Table 2: Summary of the past research**

| Year | Research Topic | Research Method | Main Research Result |
|---|---|---|---|
| 2023 | AI-augmented HRM: Antecedents, assimilation and multilevel consequences | Literature review, Framework development | Lack of in-depth research on HRM (AI), Proposed a comprehensive framework |
| 2023 | AI-assisted HRM: Towards an extended strategic framework | Systematic review | Understanding of current knowledge, Future research directions |

| 2023 | AI in HRM: case study analysis. Preliminary research | Systematic review, Bibliometric analysis | Managerial capabilities for AI adoption, Common areas for AI use |
|---|---|---|---|
| 2024 | AI Applied to HRM: A Bibliometric Analysis | Bibliometric analysis | The study revealed a significant increase in research on AI applications in HRM |
| 2024 | Application of AI in HRM: a conceptual framework | Exploratory research | AI technologies in HRM, Advantages and difficulties of AI integration |
| 2024 | AI and Automation in HRD: A Systematic Review | Author citation and co-citation analysis, Communication analysis, Exponential random graph model | Key findings include how AI and automation influence specific HRD processes and their overall effects on HRD functions. |

### *Innovations in Methodology Compared to Existing Research*

This study introduces several key innovations in the analysis of AI in HRM, which set it apart from previous works in the field:

- **Integration of Social Network Analysis (SNA) with TOPSIS for Research Collaboration Mapping**
  Previous studies have largely relied on bibliometric methods to analyze the AI-HRM research landscape. This study innovatively combines Social Network Analysis (SNA) with the TOPSIS decision-making method, providing a dual approach that not only identifies the structural patterns in research collaborations but also ranks institutions and authors based on their research influence. This combination allows for a deeper, more nuanced understanding of research dynamics compared to traditional bibliometric methods.
- **Global Mapping of AI-HRM Research Networks**
  While many studies examine AI in HRM from a regional or thematic perspective, this research offers a global mapping of research collaborations. By analyzing the co-authorship networks across different regions and countries, the study reveals global research hubs and trends that have not been fully explored in the existing literature. This global perspective is a significant methodological advancement, offering insights into international cooperation in AI-driven HRM research.
- **Identification of Key Research Themes Using SNA**
  Another methodological innovation is the use of Social Network Analysis to identify emerging research themes. While traditional methods focus on literature reviews and thematic analysis, this study applies SNA to uncover hidden patterns in the research community, identifying themes based on collaboration networks rather than just keywords or citations. This approach reveals four primary themes of AI in HRM, offering a comprehensive view of the field's evolution and potential future directions.
- **Enhanced Quantitative Analysis with Multiple Centrality Measures**
  Unlike previous studies that typically focus on one or two centrality metrics, this research employs multiple centrality measures (degree, betweenness, closeness, and eigenvector) to assess the influence of authors and institutions. This approach provides a more comprehensive and granular understanding of the research landscape, offering insights into not only who the most influential authors are but also how they contribute to shaping the field.
- **Comprehensive Dataset and Cross-Method Integration**
  By integrating a large dataset of over 100,000 authors and 287,000 collaborations, and combining quantitative methods like SNA with decision-making tools such as TOPSIS, this study offers a unique methodological framework. It enables a more accurate and holistic analysis of AI-HRM research trends than previous studies that have focused on smaller datasets or singular methods.

In summary, the methodological innovations introduced in this study—such as the integration of SNA with TOPSIS, the global mapping of research networks, the identification of emerging research themes using SNA, and the use of multiple centrality measures—mark significant advances in the field. These innovations provide a deeper and more nuanced understanding of the AI-HRM research landscape compared to previous works that primarily relied on bibliometric approaches or were limited in scope. The combination of these methods enables a more comprehensive, multi-dimensional analysis of research collaborations, offering valuable insights into the evolution and future directions of AI applications in HRM.

## III. METHODOLOGY

This scholarly article investigates the phenomenon of co-authorship within the intersection of AI and HRM. The research commenced by systematically gathering articles related to "AI applications in HRM" using a carefully curated set of keywords specific to AI and HRM [21]. The selection process involved the use of two distinct groups of keywords: one set focusing on AI-related terms and another on HRM-related terms, ensuring a comprehensive search across both domains [46]. These keywords are detailed in Figure 4 of the manuscript.

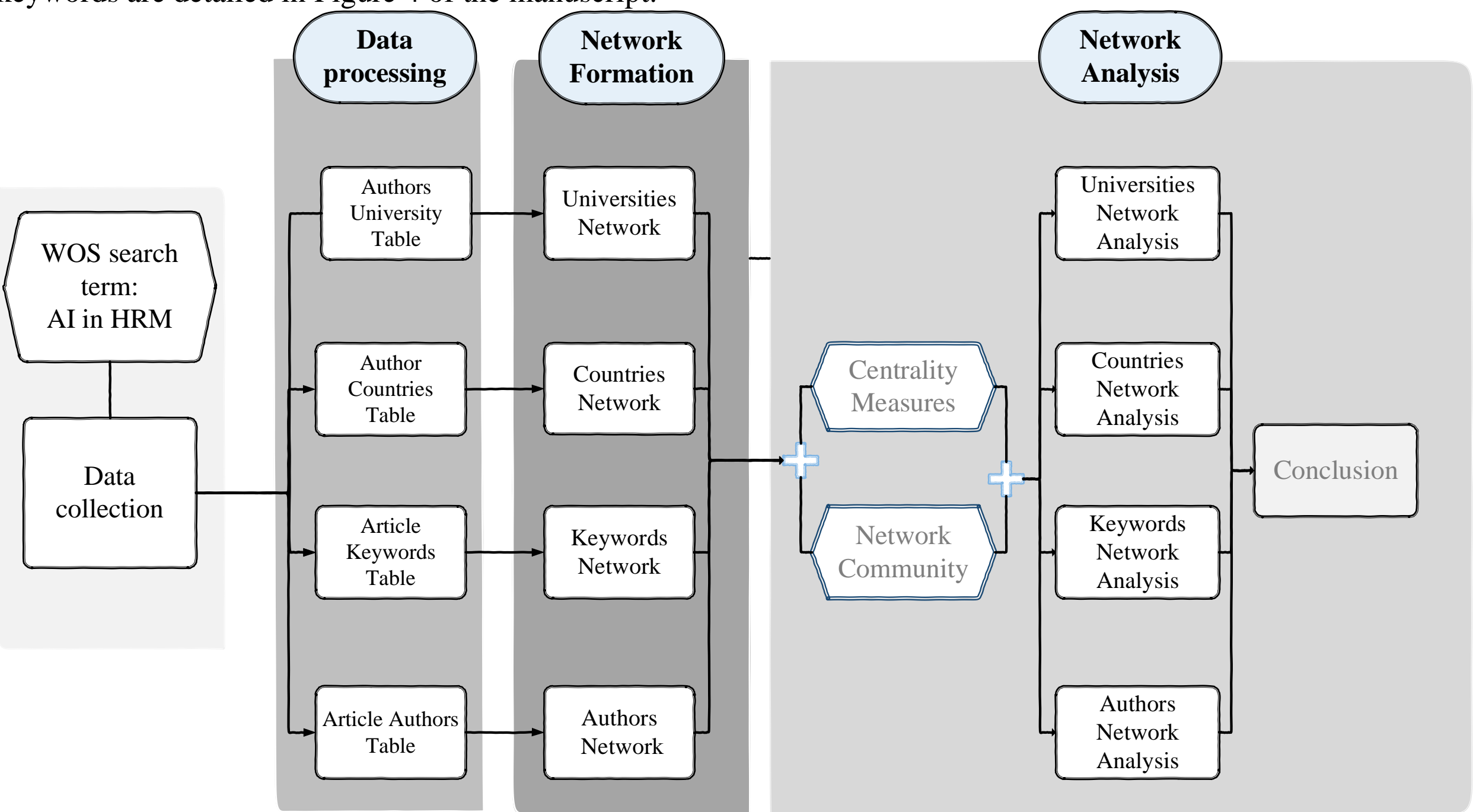

Figure 2 Methodological Framework

A strategic decision was made to utilize the Web of Science (WOS) as the primary database for this study. WOS was chosen due to its extensive coverage of high-quality, peer-reviewed research across disciplines such as business, computer science, and management [47]. The rigorous inclusion criteria, extensive citation indexing, and global reach of WOS provide a robust foundation for analyzing the complex relationships between AI and HRM [39]. By employing WOS, the research accesses a rich dataset that facilitates detailed analysis and the identification of significant trends and patterns within the field [21]. Many researchers have utilized this source for examining and analyzing co-authorship networks, as evidenced by studies such as [48-52].

Furthermore, the study predominantly relies on the Web of Science Core Collection, a prestigious resource recognized for cataloging significant scientific inquiries and widely respected in academic circles for its reliability and depth [53]. The research methodology, including the procedural steps for data collection and analysis, is thoroughly outlined in Figure 2 of the manuscript, ensuring transparency and replicability.

In the realm of data preprocessing, a dataset is formulated to facilitate the examination of the co-authorship network involving authors, universities, countries, and keywords. The process of dataset creation is illustrated in Figure 3 showcasing the establishment of links between authors when they collaborate on an article, with the weight of these links escalating as authors collaborate on multiple articles.

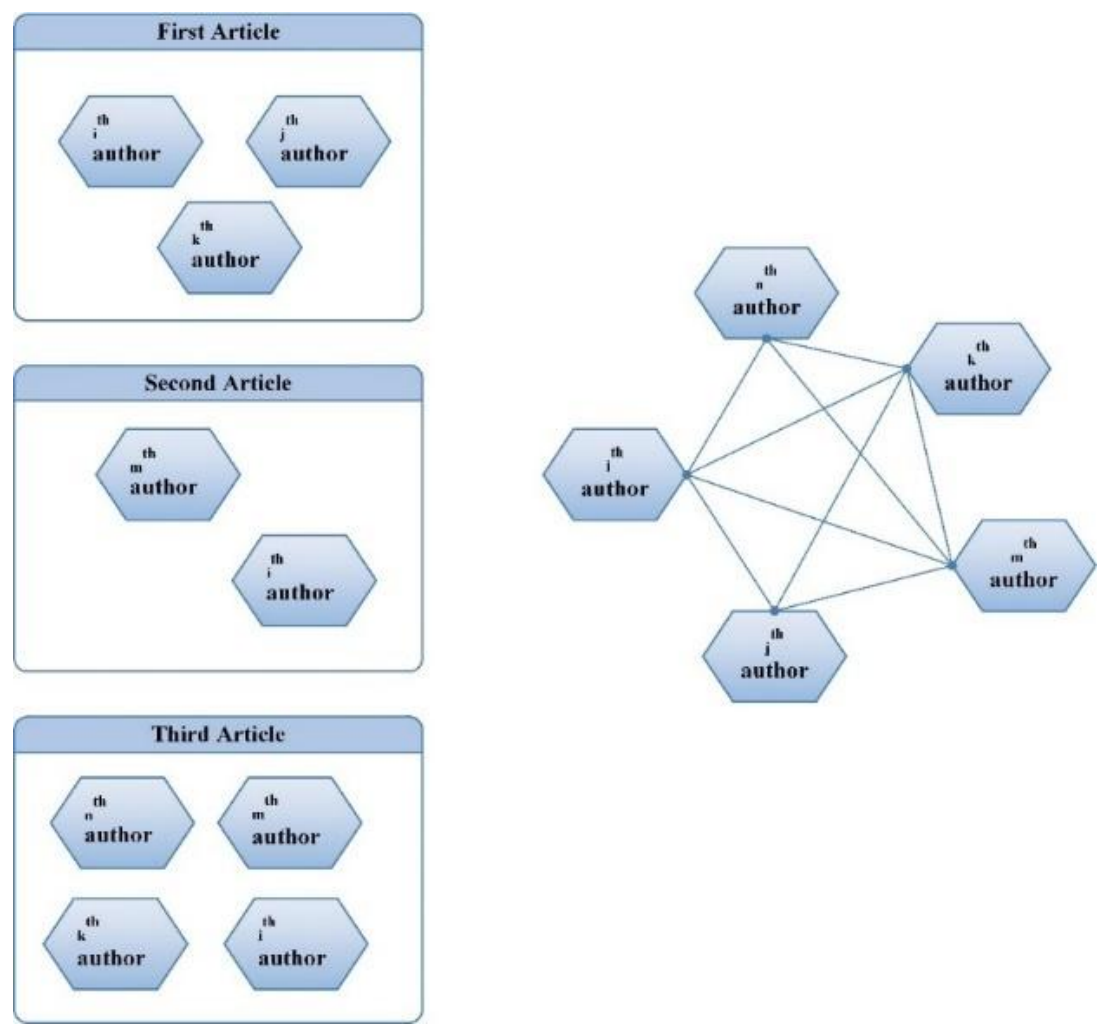

Figure 3 **Network formation** rules [54]

### *Social Network Analysis (SNA)*

Analyzing co-authorship networks is now a crucial method for examining how scholars collaborate with each other. Through the utilization of Social Network Analysis (SNA) methods, scholars can acquire more profound understandings of the formation and evolution of academic collaborations. This method is especially important in the current research environment, where it is vital to grasp new trends, recognize influential experts, and forecast upcoming research directions [55]. These insights are critical for grasping how knowledge is disseminated, how research communities form, and for pinpointing potential collaborators [56].

SNA further allows for the quantification of collaborative intensity, identification of central figures, and evaluation of overall connectivity within a research network. Visualization of these networks provides a comprehensive view of the research landscape, helping scholars to identify new trends and potential opportunities for collaboration [57, 58].

Social Network Analysis, first created in the mid-1900s, grew in popularity thanks to Stanley Milgram's "small world" experiments. Since gaining popularity through the "Six Degrees of Kevin Bacon" game, social network analysis (SNA) has developed into a flexible approach applied in various industries including healthcare, marketing, national security, political science, and particularly, research collaboration studies [59]. Essentially, SNA focuses on examining the connections between elements, depicted as nodes and edges in a network graph. This method enables researchers to discover important figures, reveal obscured connections, and analyze the distribution of resources or information in a system. SNA is a valuable method for comprehending intricate social occurrences, from illness spread to corporate conduct [60, 61].

Besides recognizing network elements, social network analysis provides numerical measurements to assess network characteristics. Indicators like density, clustering coefficient, and centrality offer important information about the unity, interconnection, and importance of nodes in a network [62]. These metrics enable researchers to compare networks, detect structural differences, and draw meaningful conclusions about the systems under investigation. In research collaboration, SNA helps map the co-authorship landscape, revealing key scholars, emerging research topics, and potential collaborators. Moreover, SNA is instrumental in uncovering hidden knowledge networks and fostering interdisciplinary collaboration[63].

Additional metrics such as community detection, density, and clustering coefficient provide further clarity into the overall characteristics of a network [64]. These metrics help in identifying cohesive subgroups, assessing overall connectedness, and understanding the distribution of collaboration patterns within the field of research. By applying these SNA techniques, researchers can better understand collaboration dynamics, pinpoint emerging trends, and develop strategies for knowledge dissemination and impact [65].

Social Network Analysis (SNA) plays a pivotal role in examining network structures through a wide array of metrics designed to quantify relationships between entities, evaluate the centrality of key actors, and identify cohesive subgroups within the network [66]. The following metrics are integral for researchers seeking to understand network dynamics and inform subsequent analyses [67, 68]:

- **Degree**: The number of connections a node has is referred to as its degree, representing the size of node iii's neighborhood. A high degree of centrality is typically observed in active individuals who maintain numerous connections with other nodes within the network. This reflects a node's influence and centrality within the system [69, 70].
- **Closeness**: Closeness centrality measures a node's proximity to other nodes in the network by calculating the average distance to every other node [71]. A node with high closeness centrality is positioned near other nodes with shorter paths, indicating its capacity to quickly access or disseminate information across the network. In a literary context, authors with higher closeness centrality tend to be more closely connected to other authors, signaling higher visibility and influence within the scholarly community [72].

- **Betweenness Centrality**: Betweenness centrality quantifies the frequency with which a node lies on the shortest path between any two nodes in the network [73, 74]. Nodes with high betweenness centrality often act as critical intermediaries or bridges, playing a pivotal role in controlling the flow of information and content throughout the network.
- **Eigenvector Centrality**: Research suggests that the top eigenvector of an adjacency matrix can be an effective indicator of network centrality [54]. Unlike degree centrality, eigenvector centrality assigns varying weights to nodes based on their connections, emphasizing the importance of influential neighbors. Eigenvector centrality provides a weighted combination of direct and indirect links, capturing the influence of both short and long-range connections within the network [75].
- **Density**: The density of a network with nnn nodes is determined by the ratio of the actual number of edges (mmm) connecting these nodes to the maximum potential number of edges in the network. This measure reflects the proportion of possible connections that are currently realized, offering insights into the network's overall connectivity [76, 77].
- **Community Detection**: Identifying communities is a key aspect of Social Network Analysis (SNA), facilitating an understanding of the structure and behavior of networks. A community is a structural element that showcases the relationship between nodes, with the detection process beginning by forecasting connections between nodes to identify more densely connected clusters. This process helps researchers better understand the underlying structure and dynamics of the network, aiding in the identification of cohesive groups that exhibit strong internal ties while maintaining weaker connections with other groups[78, 79].
- **Size**: number of the network's nodes [80].

One of the most common ways to analyze a network is to look at the centrality of various nodes to identify key players, information hubs, and gatekeepers across the network. There are three types of centralities, each corresponding to a different aspect of connectivity and centrality. Degree, Betweenness, and Closeness Centrality are measures of a node's importance. Table 3 delineates several centrality indicators, including Degree, Closeness, and Betweenness centralities [81].

**Table 3 SNA Measures**

| Centrality measure | Description | Formula |
| --- | --- | --- |
| Degree | The number of links involving that node | $\sum_{I=1}^{n} a(n_i, N_j)$ |
| Closeness | The standard deviation of the length of the shortest path from one node to all other nodes | $\frac{N-1}{\sum_{I=1}^{N} D(j,i)}$ |
| Betweenness | The number of times a node is found along the shortest possible path to other nodes. | $\frac{1}{N^2} \sum_{\forall s,t \in V} \frac{N_{S,t}^{I}}{G_{S,t}}$ |
| Eigenvector | The sum of the centrality values of the nodes To which it is connected | $\frac{1}{\Lambda} \sum_{I=1}^{N} A_{I,j}, X_J$ |

The objective of the network formation phase is to establish a co-authorship network spanning authors, universities, countries, and keywords. Utilizing the datasets generated in the preceding phase, Gephi software is employed to generate and visualize the networks. Widely recognized as a prominent tool for social network analysis and visualization, the Gephi 0.9.7 software version prepares us for the subsequent scrutiny of the networks.

The fourth phase involves the examination of the developed networks. This examination encompasses the application of centrality metrics such as Eigenvector, Degree, Closeness, and Betweenness, alongside the employment of community detection techniques.

### *TOPSIS*

A method utilized to identify the optimal choices by considering various factors is referred to as multiple-criteria decision-making (MCDM). This strategy is also known as multiple-attribute decision-making (MADM) and multiple-objective decision-making (MODM), as introduced by Hwang and Yoon in 1981 [82]. TOPSIS was selected for this study due to its ability to efficiently handle multi-dimensional data and offer a comprehensive ranking framework. Unlike traditional bibliometric methods, which rely on basic metrics like citation counts or co-authorship, TOPSIS provides a multi-criteria approach for ranking the influence of authors. By integrating multiple centrality measures (such as degree, betweenness, closeness, and eigenvector centrality), it enables a more nuanced evaluation of authors based on both their quantity of collaborations and their strategic position within the research network. The procedural components of the TOPSIS technique consist of the following steps:

**Step 1**: create a decision matrix $(\mathrm{Dij})_{p\times q}$ with p rows as options (nodes) and q columns as criteria.

**Step 2**: The dimensionless matrix $L = (lij)_{p\times q}$ is created by normalizing the decision matrix:

$$l_{ij} = \frac{D_{ij}}{\sqrt{\sum_{i=1}^{p} D_{ij}^2}}$$

**Step 3**: Each criterion's relative importance is shown by a weighted matrix T.

Calculate T as follows:

$$T = (t_{ij})_{p\times q} = (w_j l_{ij})_{p\times q} \quad \forall i,j$$

**Step 4**: The worst solution $t_j^-$ and the best solution $t_j^+$ are defined as follows:

$$t_j^+ = \{(Max\ t_{ij} | i = 1,2, \dots . p. \forall_j \in J_+), (Min\ t_{ij} | i = 1,2, \dots . p. \forall_j \in J_-)\}$$

$$t_j^- = \{(Max\ t_{ij} | i = 1,2, \dots . p. \forall_j \in J_-), (Min\ t_{ij} | i = 1,2, \dots . p. \forall_j \in J_+)\}$$

$J^+$ represents the set of benefit criteria and $J^-$ represents the set of cost criteria. A set of criteria, $J^+$, is aimed to be greater, while a set of criteria, $J^-$, is aimed to be less.

**Step 5:** Distance all alternatives (options) and the best solution $t_j^+$ is calculated as follows:

$$S_i^+ = \sqrt{\sum_{i=1}^{q} (t_{ij} - t_j^+)^2}\ , i = 1,2, \dots , p$$

Also, the distance between the whole alternatives and the worst-defined solution $t_j^-$ is calculated as follows:

$$S_i^- = \sqrt{\sum_{i=1}^{q} (t_{ij} - t_j^-)^2}\ , i = 1,2, \dots , p$$

**Step 6**: Defining the comparative closeness to the best solution as follows:

$$C_i = \frac{S_i^-}{S_i^- + S_i^+}$$

**Step 7**: Alternatives are ranked according to $C_i$

## IV. RESULT

### A. DATA GATHERING

The investigation collected a significant amount of information on the role of AI in HRM studies. To specifically identify articles related to “AI in HRM,” the search was initiated using a carefully selected set of keyword search terms, as illustrated in Figure 4. These terms cover topics pertinent to HRM alongside those relevant to AI. The choice of the Web of Science (WOS) database was based on its extensive coverage and advanced search capabilities, which are particularly well-suited for academic research. WOS enables the application of search filters to specific fields such as the title, abstract, and keywords. This functionality is crucial for refining search results and minimizing the inclusion of irrelevant articles, thereby improving the focus and relevance of the retrieved articles. Many researchers have extensively utilized WOS in their studies, highlighting its value in academic investigations across various disciplines [50, 51, 83, 84]. The detailed keyword strategy and the types of extracted documents are depicted in Figure 4.

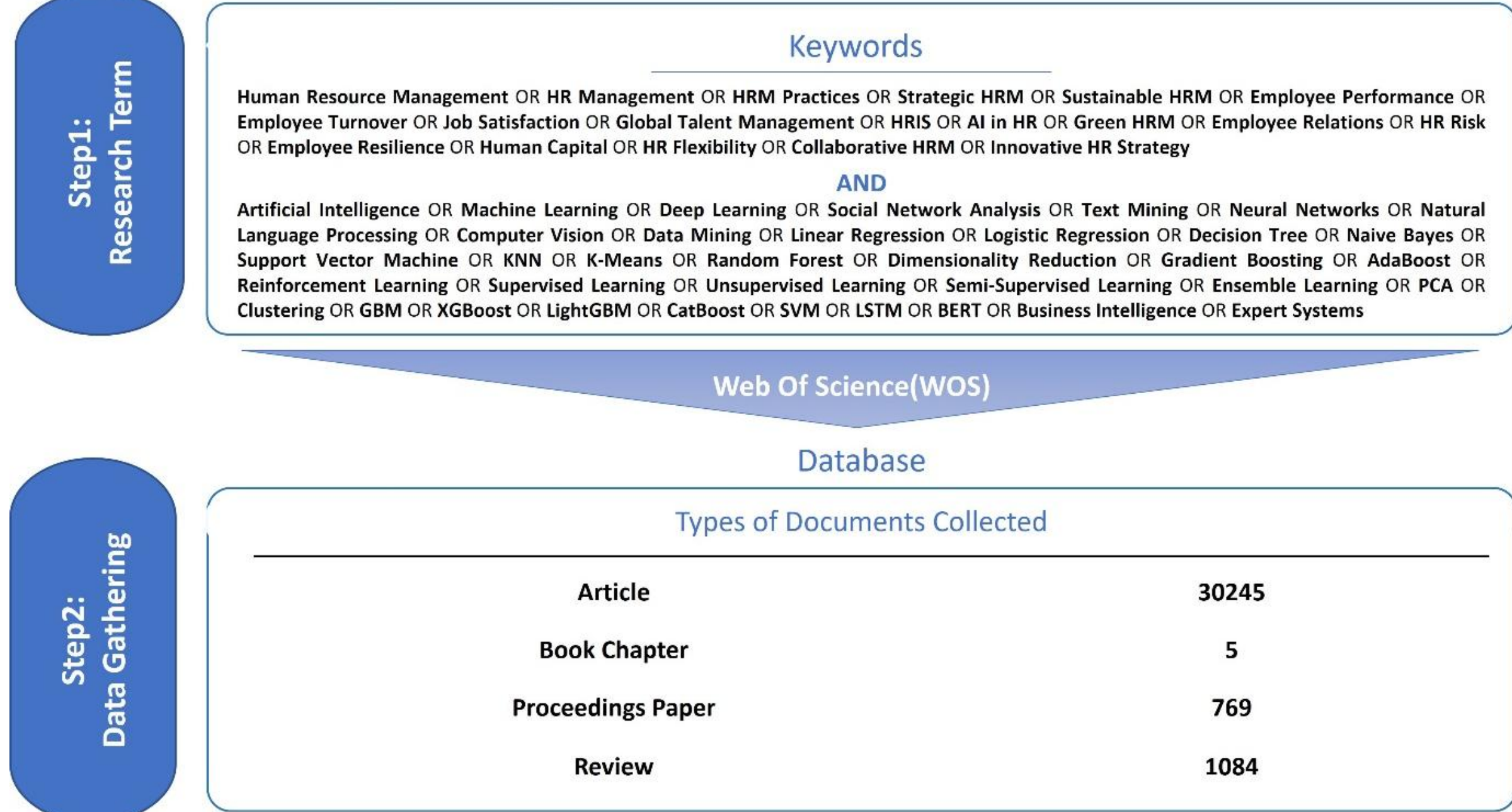

Figure 4 Search strategy and results

The initial exploration resulted in 32,103 publications based on the selected keywords. Subsequently, the team of researchers scrutinized the distribution of these publications based on their respective publication years and illustrated the outcomes in Figure 5. The data indicates a steady rise in the quantity of publications over time, reflecting a heightened interest in the subject among scholars. The complete dataset is accessible via the link provided1.

[1] https://zaya.io/AIinHRM

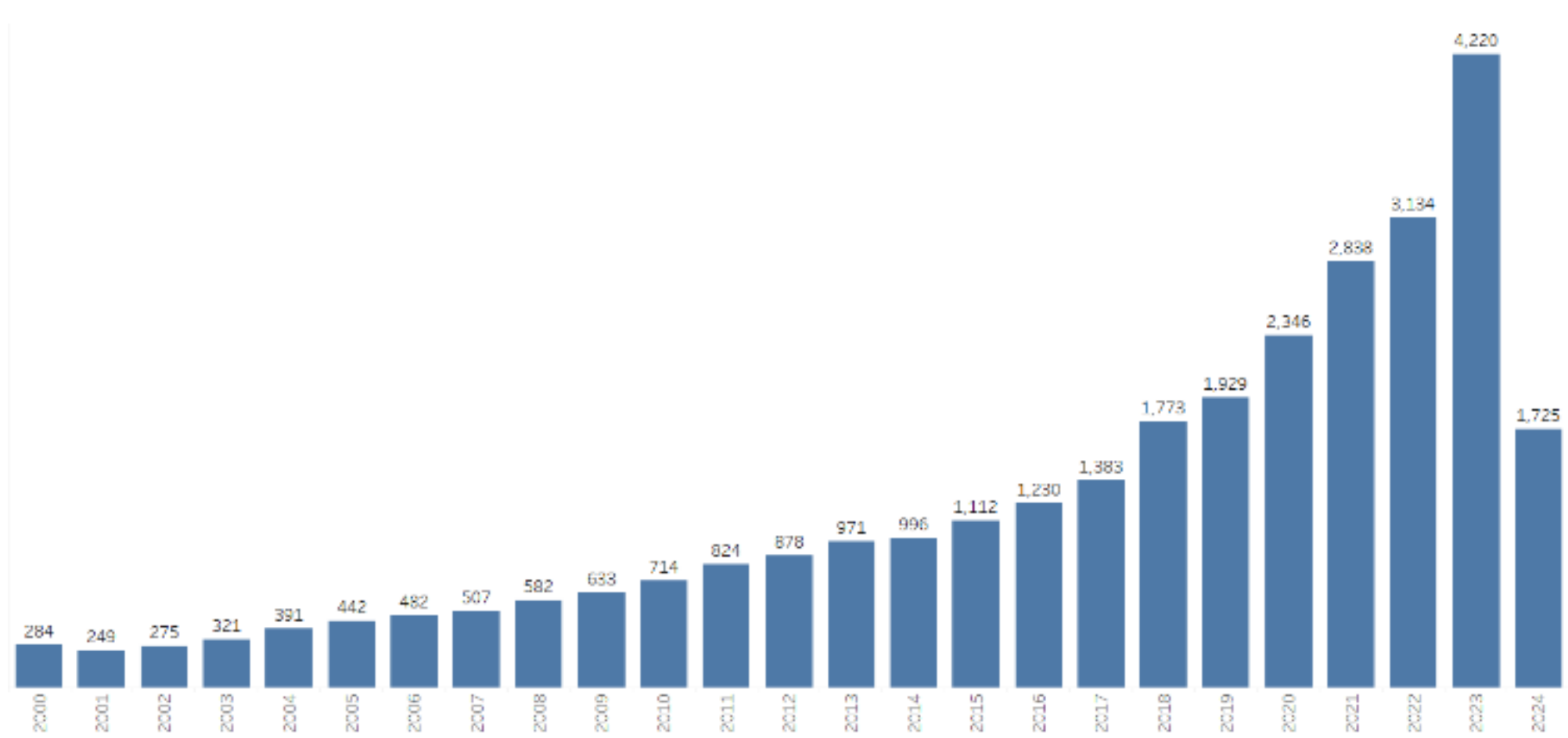

Figure 5 Year of publication

### *B. AUTHORS NETWORK*

Recognizing the important participants in the co-authorship network is essential for understanding their notable role in scientific productivity and research cooperation within the network. According to [81], these individuals are instrumental in the network's general research efforts. Figure 6 provides a visual representation of the co-authorship network, where the darker, denser central area highlights the core group of researchers with the most connections. The lighter, sparser outer areas show fewer connected nodes, possibly belonging to different clusters or communities within the network. This graph represents a social network to a biological network depending on the field of AI in HRM.

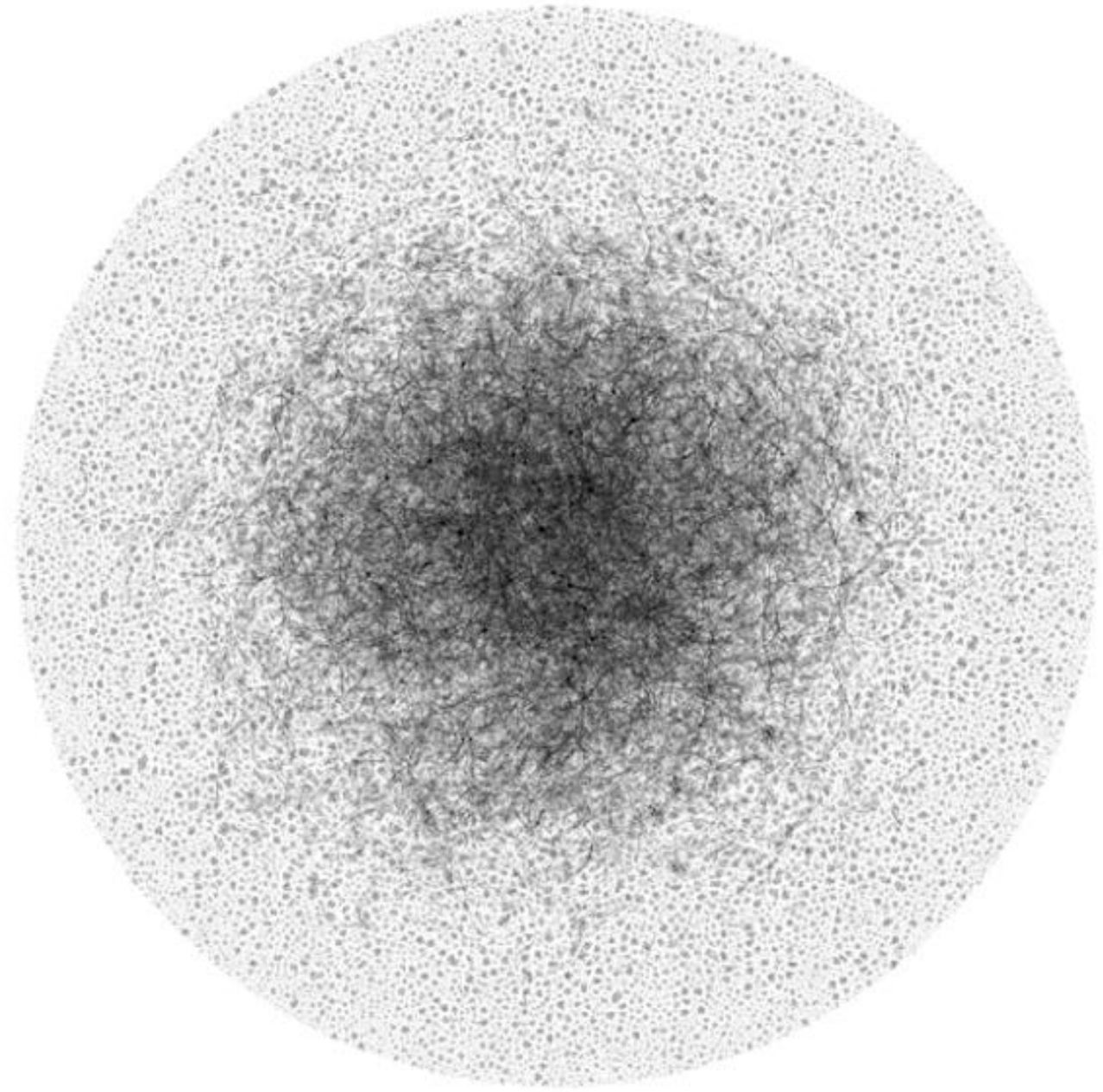

Figure 6 AI in HRM Co-authorship Network

Table 4 provides a detailed overview of the co-authorship network in the field of AI in HRM. This network analysis offers valuable insights into the structure and dynamics of collaboration within this domain.

Figure 6 consists of 102,296 individual entities (nodes), each representing an author in the AI in HRM field. With 287,799 edges, there are nearly three times as many connections as there are nodes. This high ratio of edges to nodes indicates significant collaboration among researchers in this domain. The average degree of 5.627 suggests that each author typically collaborates with about 5 or 6 others, pointing to a tendency for small to medium-sized research groups in this field.

The network diameter of 25 indicates that the furthest two authors are separated by 25 steps or connections. This considerable diameter may reflect the existence of multiple specialized sub-domains within AI in HRM that might have limited interaction.

The graph density of 0.0001 shows a very sparse network, meaning many potential connections between authors in this field do not exist. This sparsity could indicate opportunities for increased collaboration in the future.

The presence of 14,211 connected components suggests the network comprises many small, isolated clusters of authors. This structure could indicate specialization in various sub-areas of AI in HRM. The high average clustering coefficient of 0.956 further supports this interpretation, indicating that authors tend to form tightly-knit groups or communities. This suggests strong collaboration within specific research teams in the AI in HRM domain.

The dense black center of Figure 6 likely represents a highly connected core where leading and prolific authors in AI in HRM are closely linked to one another. This central group of researchers potentially plays a key role in advancing the field and bridging different research communities.

Table 4 Details the Co-authorship Network

| Metric | Value |
| --- | --- |
| Nodes | **102,296** |
| Edges | 287,799 |
| Average Degree | 5.627 |
| Network Diameter | 25 |
| Graph Density | 0.0001 |
| Connected Components | 14,211 |
| Avg. Clustering Coefficient | 0.956 |

The collaborative practices of individual authors in the field of AI in HRM were assessed using centrality measures to identify those with the highest popularity, centrality, and prominence in the co-authorship network. The authors with the most connections to other authors in the network were identified using centrality degree, and

Table 5Table 5 suggests amalgamating centrality metrics using TOPSIS, a technique for multi-attribute decision-making [81] and the authors with the highest centrality in AI in HRM. This can be utilized to measure the extent of trust and the level of collaboration among authors. Pedrycz Witold and Chen, C.L. Philip, had the most robust connections, suggesting that they collaborated most often on these publications.

Pedrycz Witold has been identified as the author who collaborated the most with others in the field of AI in HRM, with Chen, C.L. Philip, and Herrera Francisco following closely behind. These writers were seen as the most powerful and involved partners in the network, having the greatest number of connections to other authors in the AI in HRM field due to their high degree.

Table 5 Most Central Authors

| Rank | Degree | Closeness | Betweenness | Eigenvector |
| --- | --- | --- | --- | --- |
| 1 | Pedrycz Witold | Chang, Wb | Pedrycz Witold | Pedrycz Witold |
| 2 | Chen, C.L.Philip | Juang, Jg | Chen, C.L.Philip | Chen, C.L.Philip |
| 3 | Herrera Francisco | Chang, Kia-Yi | Wang, Wei | Wang, Wei |
| 4 | Wang, Wei | Tai, Wei-Shen | Herrera Francisco | He, Wei |
| 5 | Herrera-Viedma, Enrique | Chang, HH | Ding, Weiping | Liu, Yang |
| 6 | He, Wei | Yalcin, Haydar | Yang, Jie | Yang, Chenguang |
| 7 | Wang, Jing | Aldhaban, Fahad | Liu, Yang | Su, Chun-Yi |
| 8 | Liu, Yang | Melovic, Boban | He, Wei | Herrera-Viedma, Enrique |
| 9 | Shi, Peng | Harmon, Robert | Martinez, Luis | Shi, Peng |
| 10 | Wang, Zidong | Cirovic, Dragana | Yao, Xin | Wang, Shitong |
| 11 | Nie, Feiping | Basoglu, Nuri | Wang, Zidong | Shi, Peng |
| 12 | Li, Tianrui | Veljkovic, Slavica Mitrovic | Wang, Yu | Wang, Zidong |
| 13 | Wang, Yu | Vulic, Tamara Backovic | Shi, Peng | Lin, Jerry Chun-Wei |
| **14** | Wang, Shitong | Alassaf, Deemah | Yang, Bo | Herrera, Francisco |
| **15** | Wang, Jun | Ahrens, Diane | Li, Tianrui | Wang, Shitong |
| **16** | Li, Hui | Yucelbas, Cuneyt | Hu, Qinghua | Wang, Jun |
| **17** | Wang, Hao | Shifter, Dara | Chen, Hao | Wu, Min |
| **18** | Jiao, Licheng | Wohlrab, Lars | Kaynak, Okyay | Wang, Yu |
| **19** | Li, Yan | Smirnov, Evgueni N. | Yang, Chenguang | Yun, Unil |
| **20** | Hong, Tzung-Pei | Sprinkhuizen-Kuyper, Ida G. | Wang, Jun | Chen, Long |

Authors with a high betweenness centrality are seen as vital in linking various author groups, as per social network analysis. This promotes interaction and cooperation among typically unconnected groups. Pedrycz Witold had the top betweenness centrality in the AI in HRM co-authorship network, functioning as an intermediary or connector. Chang, Wb possessed the greatest closeness centrality, demonstrating substantial impact within the network. Writers such as Chen, C.L. Phillip, and Wang, Wei played significant roles in the network as well. When new AI subjects or uses are introduced, they are often imitated by other authors such as Chang, Kia-Yi, and Juang, Jg.

Pedrycz Witold served as a facilitator in the collaboration network of AI in HRM. His ties to the other writers were restricted. In Table 6, Pedrycz Witold showed the highest eigenvector centrality, indicating he is the most influential person in the network and likely collaborated with other important individuals.

Selecting the crucial and influential nodes poses a challenge due to the abundance of metrics and outcomes to consider.

**Table 6 20 effective authors**

| Rank | Author | Institute | Country | Field of study |
|---|---|---|---|---|
| 1 | Pedrycz, Witold | University Of Alberta | Canada | Faculty Of Engineering - Electrical & Computer Engineering Dept. |
| 2 | Chen, C. L. Philip | South China University of Technology | China | Computational, Intelligence Intelligent, Control Cybernetics, Intelligent Transportation Systems, Data Science |
| 3 | Wang, Wei | University of California, Los Angeles | USA | Data Mining, Machine Learning, Big Data Analytics, Bioinformatics, and Computational Biology Computational Medicine |
| 4 | Herrera, Francisco | Granada University | Spain | AI, Computational Intelligence, Data Mining, Evolutionary Algorithms, Big Data Analytics |
| 5 | He, Wei | Purdue University Northwest | Indian | Strategic Inter-Firm Relationship Management, Global Innovation, and Knowledge Management |
| 6 | Liu, Yang | Nanyang Technological University | Singapore | Cybersecurity, Software Engineering and AI |
| 7 | Wang, Zidong | Brunel University London | England | Intelligent Data Analysis, Control Engineering, Signal Processing, Bioinformatics |
| 8 | Yao, Xin | University of Birmingham | England | Evolutionary Computation, Neural Network Ensembles, Multiple Classifiers, Meta-Heuristic Algorithms, Data Mining, Global Optimization |
| 9 | Shi, Peng | The University of Adelaide | Australia | Systems And Control, Electrical And Electronic, Engineering, Cybernetics, Human-Computer Interaction |

| 10 | Yang, Jie | Delft University of Technology | Netherlands | Human-Centered AI, Crowd Computing, Human-Centered Computing, Human Language Technologies, Recommender Systems |
|---|---|---|---|---|
| 11 | Su, Chun-Yi | Concordia University | Canada | On holonomic Mechanical Systems, Mechatronics, Fuzzy Control Techniques |
| 12 | Martinez, Luis | University Of Jaén | Spain | Fuzzy Decision Making, Computing with Words, Recommender Systems |
| 13 | Yang, Chenguang | University Of the West Of England Bristol | England | Neural Networks and Learning Systems, Robotics |
| 14 | Wang, Yu | University Of Hong Kong | China | Analytic Of Geo-Data, Geotechnical Risk and Reliability, Probabilistic Site Characterization |
| 15 | Ding, Weiping | Nantong University | China | Feature Extraction, Pattern Classification, Rough Set Theory, Image Segmentation, Decision Making |
| 16 | Herrera-Viedma, Enrique | University Of Granada | Spain | Fuzzy Sets, Fuzzy Decision Making, Computing with Words |
| 17 | Li, Tianrui | Southwest Jiao Tong University | China | Data Mining, Big Data, Cloud Computing |
| 18 | Wang, Jun | University College London | England | Machine Learning, Multi-Agent Learning, Information Retrieval |
| 19 | Wang, Shitong | Massachusetts Institute of Technology | USA | Nuclear Science and Engineering |
| 20 | Wang, Jing | University of Guangdong, | China | Data Mining, Big Data, and Educational Theory Technology |

Following the identification of the top 20 authors using TOPSIS, we explored the areas of expertise of these authors which are shown in

Table 6, and found that they work in data analysis, with a particular interest in HR, a trend that has been growing in recent years. This enabled us to determine the top authors' majors and areas of study.
Out of the nine countries listed in the table of the top 20 most central researchers and their countries, China is represented by five authors. England boasts four authors, Spain has three, while Canada and the United States each have two, with countries such as India, Singapore, the Netherlands, and Australia having just one representative each

Table 7 offers a thorough compilation of each country and the number of writers represented. This table displays the top countries of the most successful authors within the AI and HR field.

**Table 7: Country of 20 effective authors**

| Number of Effective Authors | 5 | 4 | 3 | 2 | 1 |
|---|---|---|---|---|---|
| Country | China | England | Spain | Canada & USA | Indiana, Singapore, the Netherlands, and Australia |

Studying the scientific connections established between various countries is a useful and intriguing analysis in the co-authorship network. The visual representation of these publications is shown in
Figure 7 illustrating the country co-authorship network.

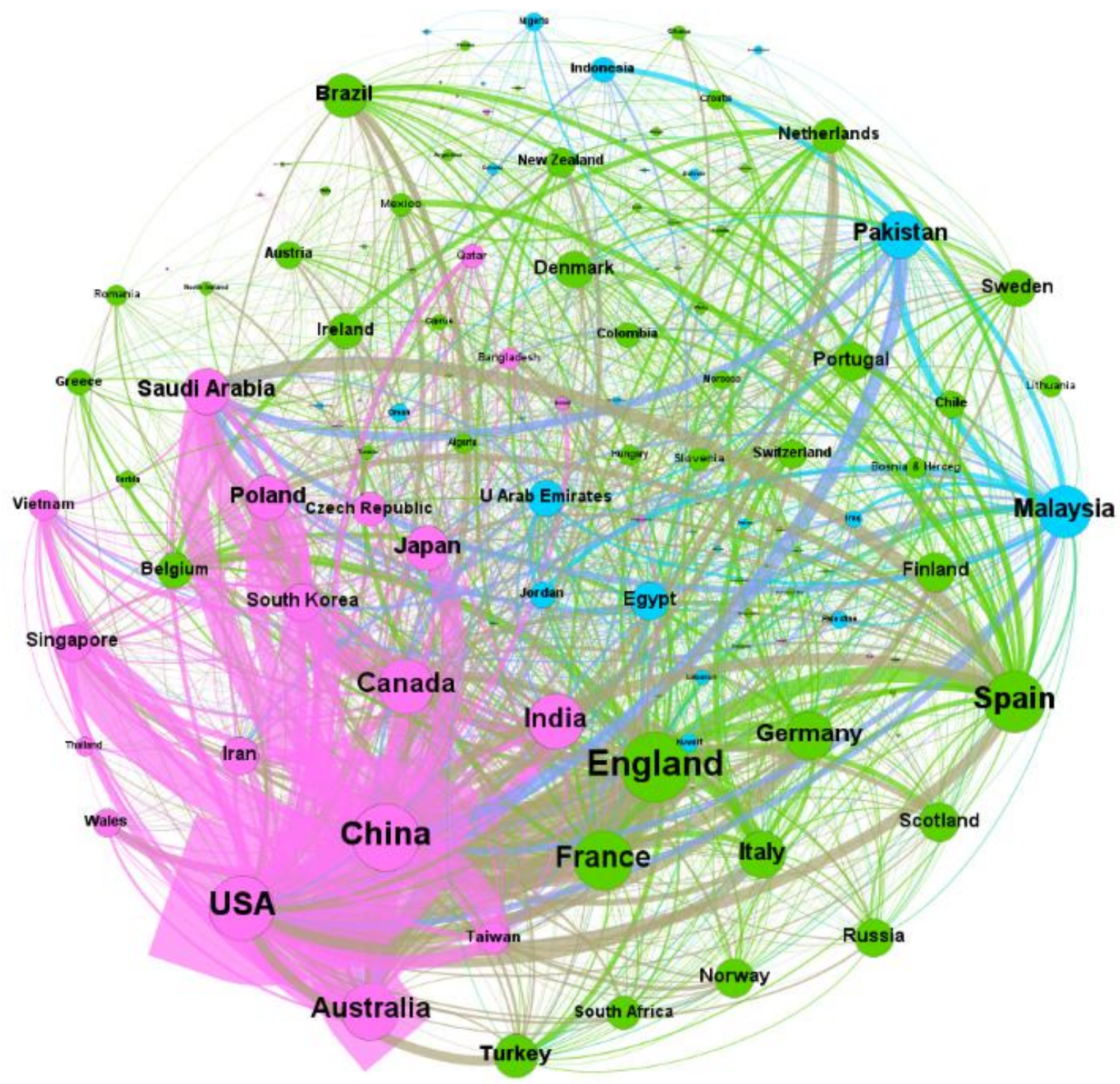

Figure 7 Co-authorship network of countries

Research on the intersection of AI and HRM has been conducted by scholars worldwide. A network analysis reveals collaborations among 127 countries through 1524 links. China and the United States are prominent collaborators, followed by Canada and China, then the United States. Leading nations in collaborations include China, the United States, Australia, England, and India. Community detection in the network shows close collaborations among countries in the field of AI in HRM research.

The green population is the largest. The green community consisted of 69 countries, with England, France, and Spain being the top three in AI in HRM research. However, the most efficient country (USA, China, Canada) is part of the Pink Community, which comprises 32 countries.

Table 8 analyzes the country's co-authorship network and includes three Communities.

Table 8 Communities of country co-authorship network

| Community color | Suggested Name | Number of edges and nodes | Top 5 country | Density |
|---|---|---|---|---|
| Green | The Global HR Applications | Nodes 69<br>Edges 475 | England, Spain, France, Germany, Italy | 0.202 |
| Blue | HRM in the Middle East and Asia | Nodes 26<br>Edges 68 | Malaysia, Pakistan, Egypt, UAE, Jordan | 0.209 |
| Pink | Global Integration of AI in HRM | Nodes 32<br>Edges 160 | China, USA, Australia, India, Canada | 0.323 |

In the expansive world of global HR, three distinct communities have emerged, bringing diverse viewpoints and valuable insights on effectively incorporating AI into HR operations. As companies adjust to the constantly changing HR environment, these communities serve as centers for creativity, collaborating to shape the future of managing employees by incorporating advanced AI technologies.

These three communities set themselves apart by focusing on different aspects:

- ***The Global HR Applications***: This community emphasizes the influence of European countries such as England, Spain, France, Germany, and Italy in the field of Global HR. The joint efforts of participants from these countries show a common commitment to incorporating AI into HR. The participation of these European nations showcases the broad global representation in the community, aiding in a comprehensive evaluation of the impact of AI on HR practices globally. As key members of the European Union, these countries influence EU decisions concerning AI and HRM, impacting the global stage with their actions.

- ***HRM in The Middle East and Asia***: This section focuses on the leading role of Malaysia, Pakistan, Egypt, the United Arab Emirates, and Jordan in implementing innovative HR practices within emerging markets. Writers from these economies are playing a significant role in influencing HR strategies, especially with the increasing presence of AI and the demand for creative HR administration. These nations need further study because of their expanding economies, with HR practices being shaped by both government rules and societal beliefs. Of great geopolitical importance, they are essential for trade and economic interactions in nearby areas. By investing in infrastructure and prioritizing skill development, these countries are leveraging their young workforce and AI to improve global business practices by evolving HR strategies.
- ***Global Integration of AI in HRM***: In countries such as China, the USA, Australia, India, Canada, Iran, Poland, Japan, Singapore, Vietnam, South Korea, and Taiwan, it is essential to have international cooperation for AI-powered workforce management. Writers from technologically advanced countries are collaborating to improve AI-based workforce management. This collaboration showcases the worldwide significance of AI in influencing contemporary workplace tactics and HR protocols. These nations give importance to diversity and inclusion in the work environment to foster inclusive atmospheres. They have crucial functions in the worldwide economy, trade, business, politics, and diplomacy. The incorporation of AI in businesses, particularly in HR management, has transformed the operational processes, enabling the creation of a diverse and inclusive workplace.

.

### C. UNIVERSITIES NETWORK

The analysis of organizational collaboration provides valuable insights into the most efficient and influential institutions [85]. The involvement of organizations in the fields of HRM and AI has laid the foundation for an institutional co-authorship network. This network, consisting of 4,036 nodes and 24,655 links, highlights collaborative efforts in publications related to HRM and AI. The size of each node represents the total number of connections among institutions, with each node signifying a distinct institution. Research partnerships are more likely to occur when there is a stronger connection between institutions. Numerous organizations have collectively contributed to papers in the fields of HRM and AI.

This institutional co-authorship network has been established through a collective effort of numerous organizations, each contributing to the advancement of knowledge in the fields of HRM and AI. The network's density and connectivity suggest that these organizations have formed meaningful relationships, collaborating on research projects and sharing knowledge and expertise. This increased collaboration has led to a significant body of research that has shed light on the impact of AI on HRM practices and the implications for organizations.

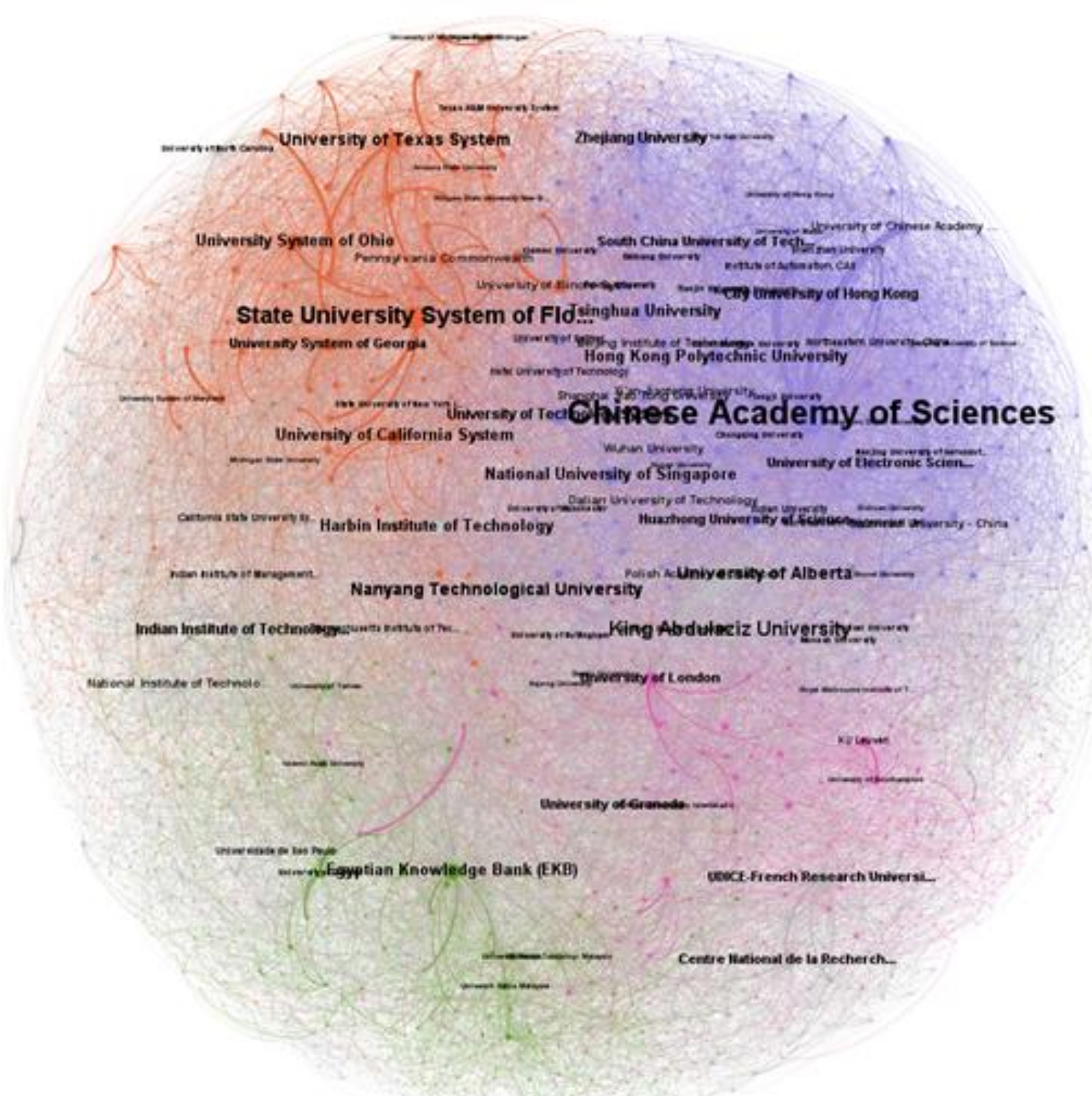

**Figure 8 AI in HRM institutional co-authorship network**

The various colors in Figure 8 represent different communities of institutions. In the graph, inter-institutional co-authorship is indicated by "edges," while each institution is represented by a "node." The thickness of the edges illustrates the level of scientific collaboration between two nodes, with thicker edges indicating stronger cooperation. A detailed overview of the five communities is presented in Table 9.

**Table 9 Details of five communities**

| Community Name | Community Color | Nodes and Edges | Density |
|---|---|---|---|
| AI Research Centers that are Dominant in Asia | Purple | Nodes 664<br>Edges 4093 | 0.015 |
| AI in leading universities in the United States | Orange | Nodes 619<br>Edges 2579 | 0.013 |
| AI in Higher Education on a Global Scale | Green | Nodes 415<br>Edges 1264 | 0.015 |
| Leaders in AI research in Europe | Pink | Nodes 361<br>Edges 981 | 0.019 |
| Newly established centers dedicated to AI research | Gray | Nodes 192<br>Edges 643 | 0.023 |

Different approaches can be used to categorize universities and scientific institutions within the AI sector, taking into account factors such as cultural and social contexts, the developmental stage of countries, and the presence of major research hubs like those in the U.S. Based on these criteria, institutions were classified and segmented into communities, with each community named according to its classification method.

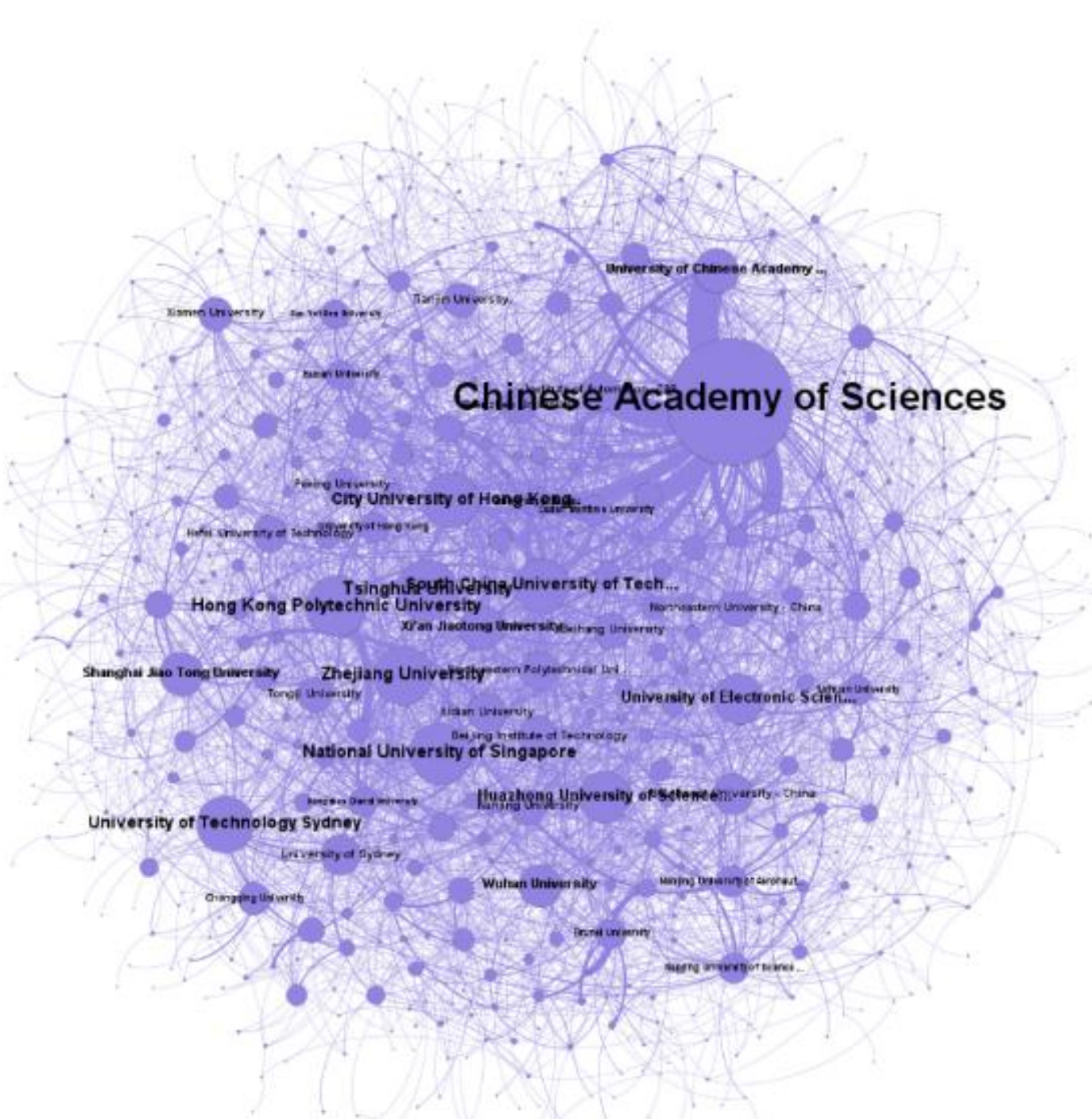

Figure 9 AI Research Centers Dominant in Asia

**AI Research Centers are Dominant in Asia**: AI Research Centers that are Dominant in Asia: Asia has emerged as a notable contributor to AI research, with the Chinese Academy of Sciences taking the lead in establishing partnerships with different research institutions in China, Hong Kong, and Singapore. The goal of this joint initiative is to further research on AI in the area [86]. Moreover, Chinese enterprises such as Lenovo, Formosa Plastics, Haier, and Huawei Technology have effectively implemented Western HRM, granting them a competitive advantage in the market [87]. The incorporation of these methods equipped them with the essential skills and knowledge for contemporary HRM practices [88]. These companies encountered difficulties in choosing and advancing staff through personal relationships with leaders, which led to the necessity of implementing competency-based selection and performance-driven incentive systems from Western HRM [89]. As Figure 9 illustrates, the adoption of these Western HRM practices has significantly impacted the performance of these Chinese companies. To remain competitive and ensure long-term growth, Chinese companies need to keep incorporating leading global AI research and HRM practices as they navigate the changing business environment [90].

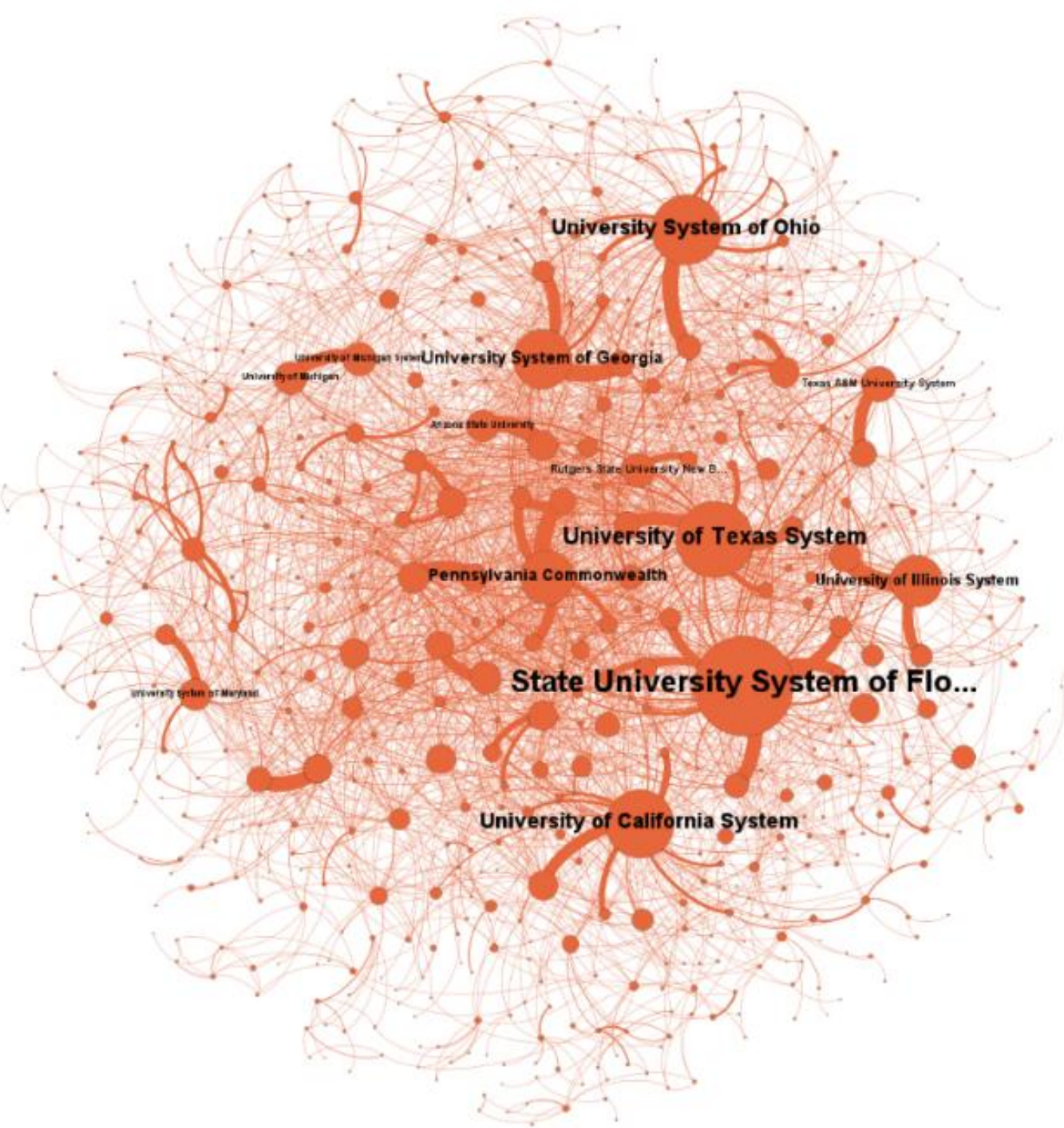

Figure 10 AI in leading universities in the United States

**AI in leading universities in the United States**: American universities are part of the vibrant orange community, which consists of esteemed institutions and systems in the United States. This network partnership involves a mutual emphasis on the importance of AI in higher education and research in the country [91]. The University of Florida, the University of Texas, and the University of California are prominent participants in this energetic community. These groups collaborate closely in the realms of AI and HRM[21].

The United States is experiencing distinct obstacles as a result of its economic success, which is causing a lack of workers. Creative hiring techniques are being implemented with a focus on varied backgrounds. HR professionals are acknowledging the necessity of effectively managing virtual teams [9]. HR professionals need to acquire technological skills to adjust to the evolving dynamics of the workplace. HR is transforming into a valuable partner, aligning its efforts with the objectives of the organization. Companies can gain a competitive advantage by implementing high-performance work systems, offering effective incentives, and maintaining a positive employer brand image to attract and retain talented employees. This strategic method plays a key role in the organization's overall achievement [92, 93].

American colleges are currently dealing with the task of integrating advancements in AI with HRM growth, with the orange group emerging as a hub of creativity driving progress in academia, research, and staff supervision [94]. Figure 10 demonstrates the rapid growth of AI adoption within the orange group universities. The competitive positioning of these institutions and businesses is made possible through the integration of AI technologies and strategic HR practices.

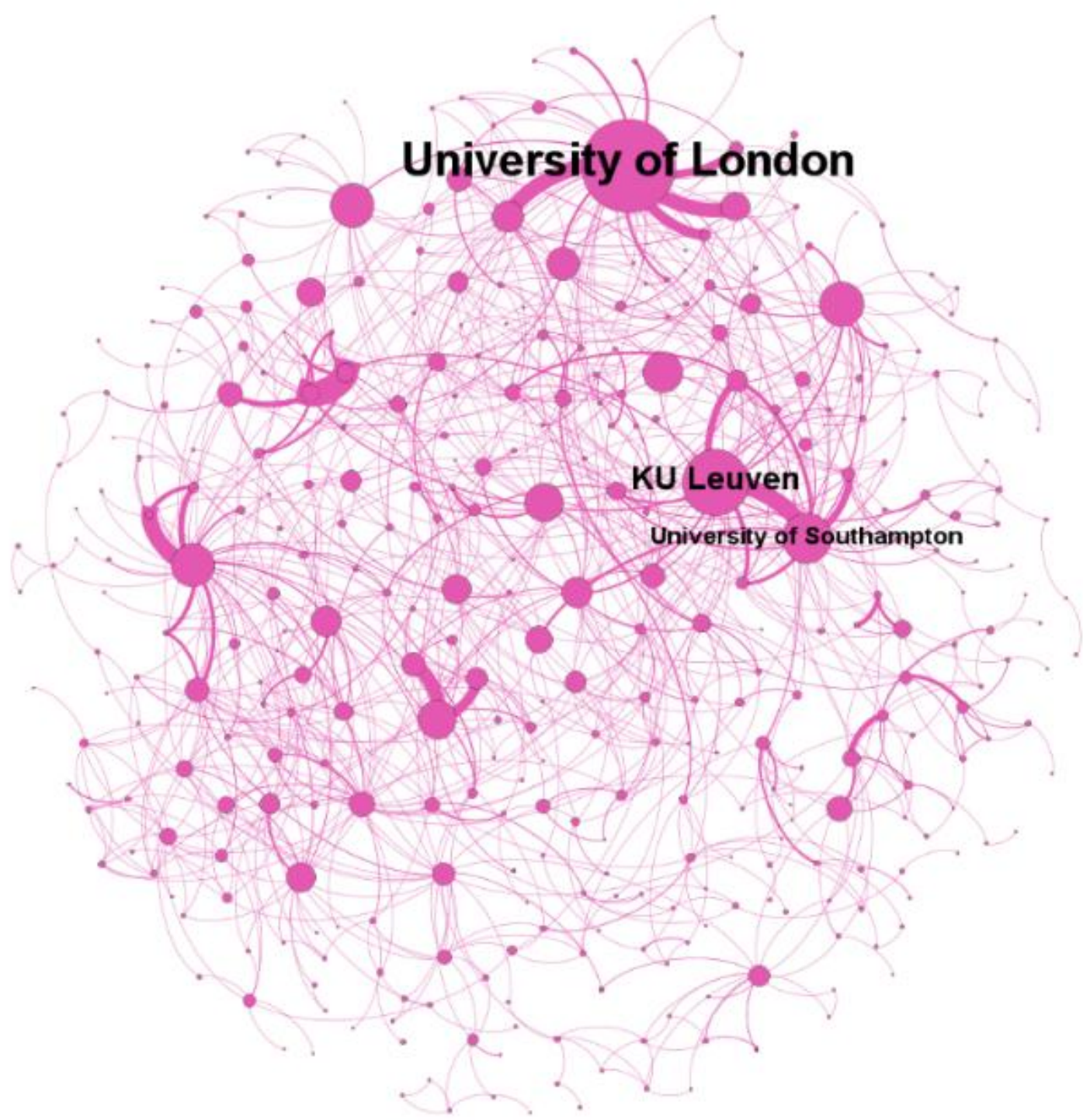

Figure 11 Leaders in AI Research in Europe

**Leaders in AI research in Europe**: Within the Pink community, a cluster of prominent European universities has emerged as leaders in cutting-edge AI research. The London University, Edinburgh University, and KU Leuven are among the premier institutions that feature prominently in Figure 11. The unique characteristics of each institution have converged to create a vibrant research ecosystem, where AI research is being pushed forward by talented researchers and scholars. HRM practices in Europe differ from those in the United States by being more informal. [17] states that European traditions consist of more relaxed grading systems and organizational structures. Workers in Europe favor a hierarchical management approach, whereas the U.S. offices typically rely on a system of organizing and reporting to oversee task completion.

A key difference between HRM in Europe and the United States lies in the management approach. European offices typically employ a hierarchical management structure, where decisions are made through a clear chain of command. In contrast, US offices often rely on a more decentralized system, where employees are given greater autonomy to complete tasks and projects, with regular reporting and feedback from their supervisors.

These differences in HRM practices have significant implications for AI research in Europe. The more relaxed and collaborative culture has fostered a sense of community and cooperation, allowing researchers to share knowledge and expertise, and to collaborate on projects that might not have been possible in a more formal environment. Additionally, the hierarchical management approach in European offices has enabled researchers to focus on specific areas of AI research, and to allocate resources effectively, rather than being bogged down in bureaucracy.

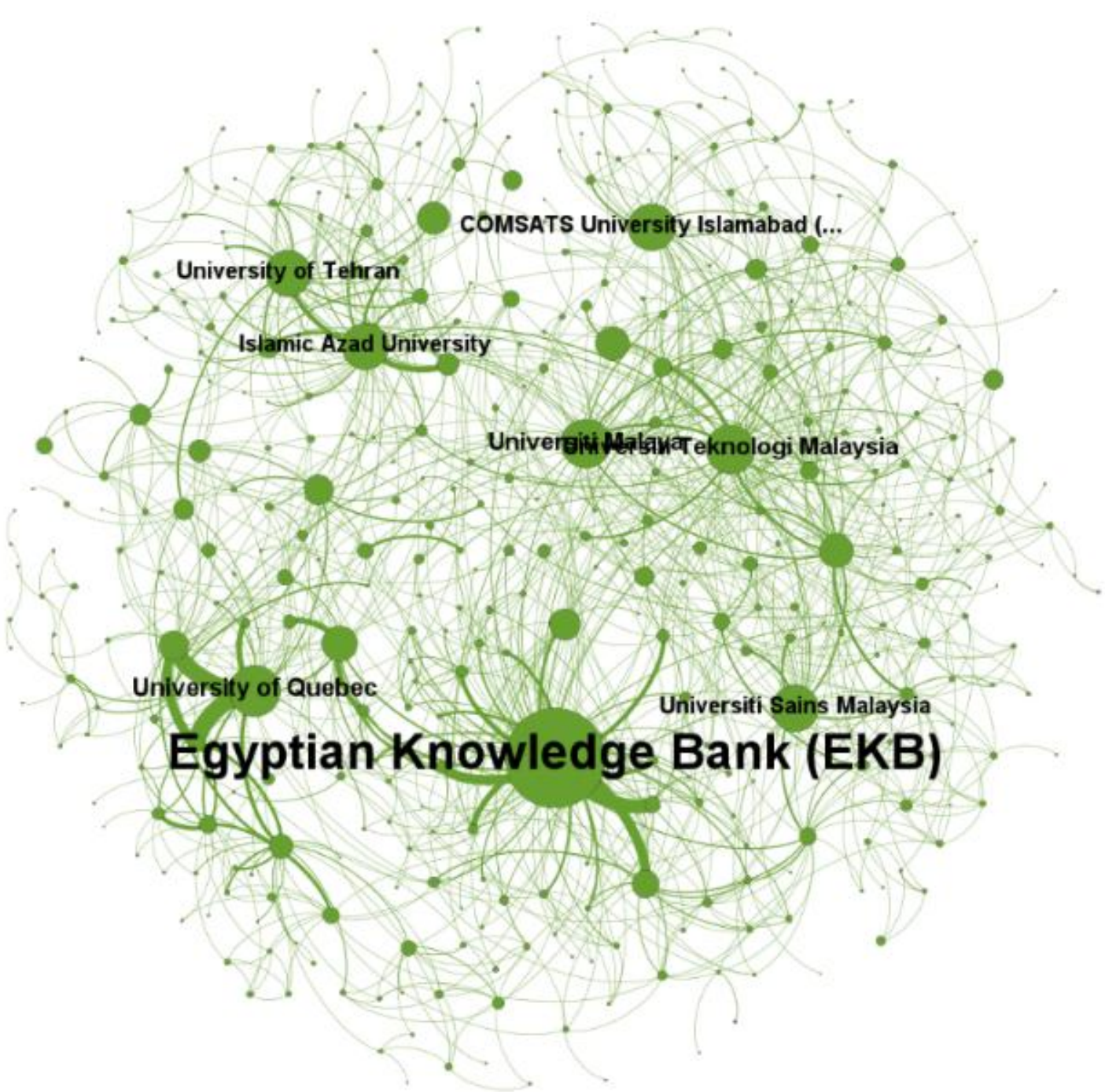

Figure 12 AI in Higher Education on a Global Scale

**AI in Higher Education on a Global Scale**: The lively green community which is shown in Figure 12 made up mostly of Iran, Egypt, and Malaysia, is actively promoting collaboration and exchanging ideas on integrating AI into academic environments. Universities from different countries unite in this diverse group to showcase a worldwide outlook on the significance of AI in higher education [21].

A recent study has emphasized the importance of HRM practices in achieving organizational success. Utilizing HRM functions effectively can serve as a strategic advantage for companies to attain success [95]. Effectively utilizing HRM functions can serve as a strategic advantage for companies seeking to attain success. The research highlights the interconnectedness of culture and HRM, with a specific focus on Islamic HRM methods in nations like Iran, Egypt, and Malaysia. Islamic principles play a crucial role in guiding conduct, but the lack of well-defined principles can create challenges when attempting to apply Islamic HRM techniques in Muslim nations. The study identifies eight essential elements of Islamic HRM and proposes that incorporating these principles can be beneficial for both companies and their staff members. Furthermore, exploring international perspectives on AI in academia involves examining cultural variations in HRM tactics, acknowledging the importance of context-dependent HRM practices [26].

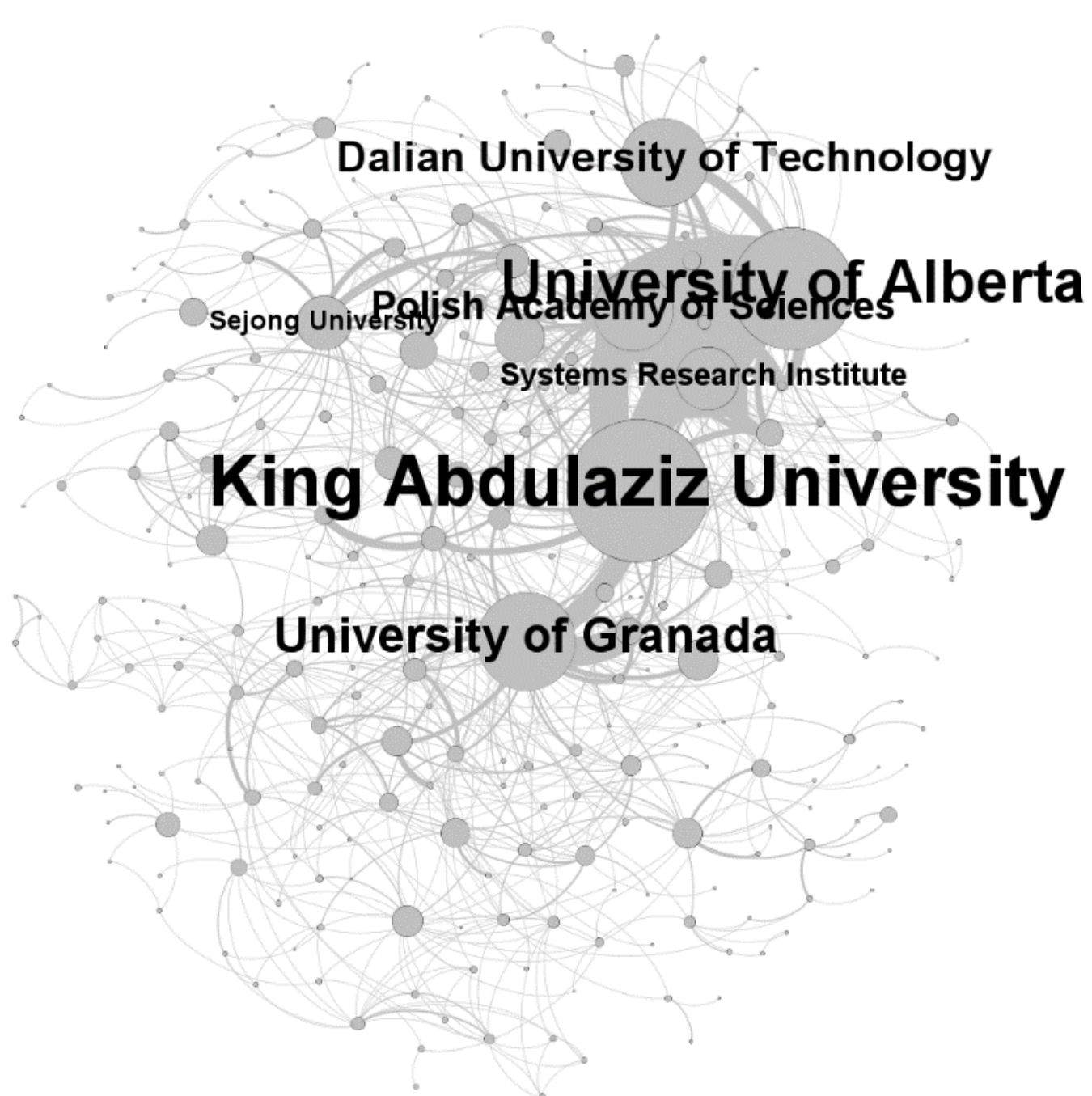

Figure 13 Newly established centers dedicated to AI research

**Newly established centers dedicated to AI research**: Figure 13 illustrates a cluster of universities and research institutions, forming a diverse group that combines developing and progressing economies. This partnership embodies a unified effort to promote progress in AI research within these fields, highlighting the potential impact of AI on the broader international AI environment. The collaboration within this group is poised to have a profound effect on the growth and advancement of AI globally, as seen in the emergence of international efforts such as the Sino-European Laboratory of Informatics, Automation and Applied Mathematics (LIAMA) [96]. Originally established to encourage Chinese colleagues to publish in international journals, LIAMA has evolved into a collaborative partnership on the global stage, marked by equal contributions from stakeholders. This is a testament to the power of international cooperation in AI research, as well as the potential for knowledge sharing and innovation. Moreover, these centers are not only advancing AI but also addressing pressing global challenges, aligning their work with the United Nations' Sustainable Development Goals (SDGs) [97]. The AI for Social Good movement, for instance, focuses on leveraging AI to drive positive change and address pressing societal issues, such as poverty, inequality, and environmental degradation. This strategic alignment enables researchers to tackle complex problems, while also promoting sustainable development and human welfare [98].

### *D. Keywords Network*

The presence of specific words within a set of articles on a given topic can reveal the primary subjects addressed in those articles. The key concepts of an article are often reflected in its keywords. Words that cluster together in documents typically relate to closely connected themes. Each cluster encompasses a different set of subject-specific keywords. By analyzing these keywords, one can discern the relationships and proximity between various topics [99].

In the context of AI & HRM research, keywords can be organized into four distinct communities, as shown in Figure 14. According to [100], keyword search analysis is a valuable method for predicting emerging scientific trends.

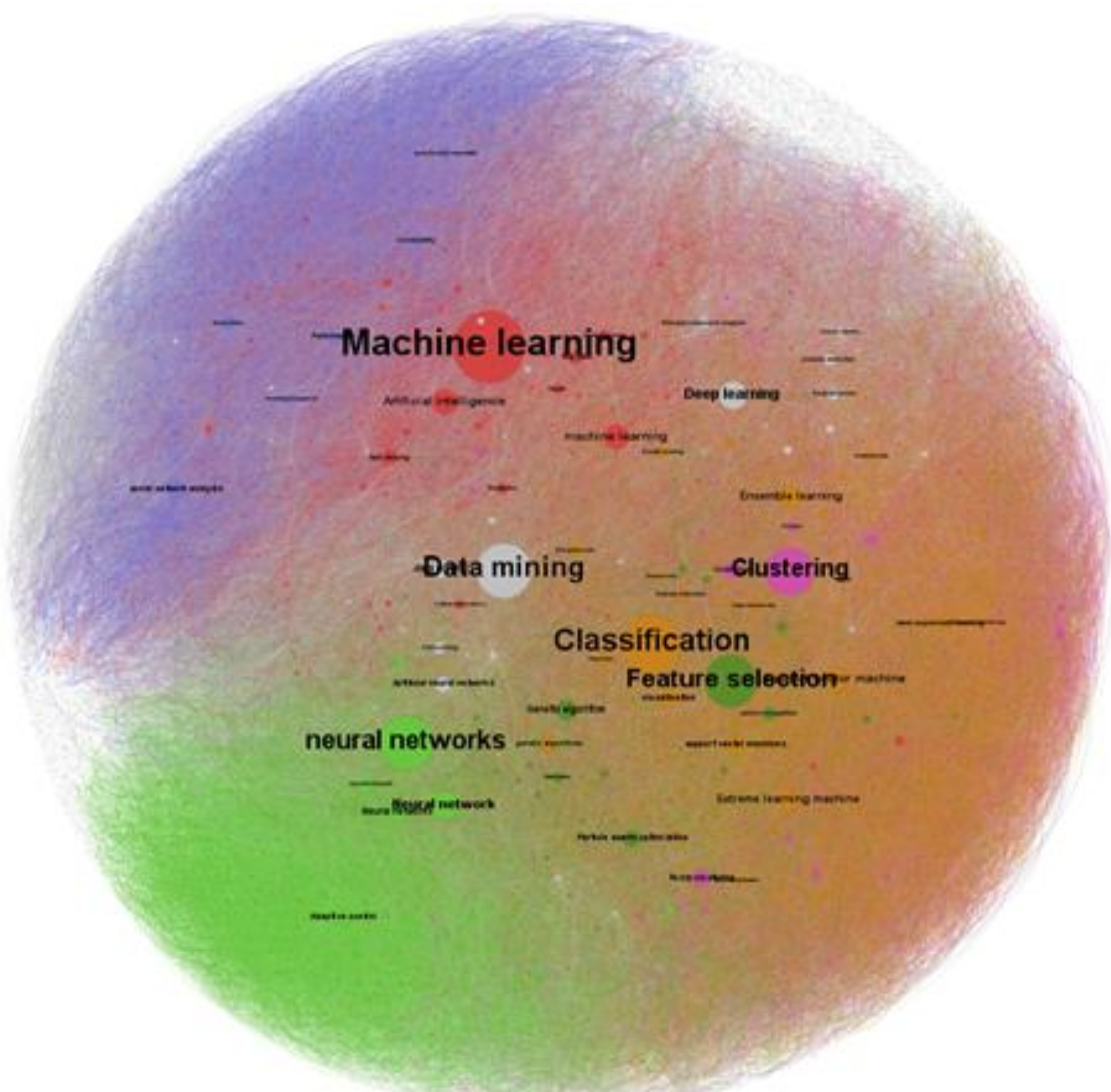

Figure 14 Keywords network

The AI & HRM articles' keywords can be categorized into four Communities according to the results of community detection. Keyword search analysis is identified by [100] as an effective strategy for forecasting upcoming scientific trends.

A network of keywords is studied, with 176894 links and 39363 nodes. The number of nodes representing each keyword shows how often it appears in the record. Keywords are crucial terms that aid readers in finding and understanding the concepts and information presented in research papers.

A closer examination of the data within communities may lead to a clearer understanding of how words are allocated to different communities and the relationships between these communities of words. Identification of four keyword communities was achieved through community detection in the indexed corpus keywords. Table 10 provides in-depth data on the number of edges and nodes.

Table 10 Details of keywords community

| Color | Number of Nodes and Edges | Density | Topics | Top 5 of Community |
|---|---|---|---|---|
| Red | Nodes 4099<br>Edges 12739 | 0.002 | AI and<br>Big Data | ML, AI, Big data, Text mining, Prediction |
| Purple | Nodes 5734<br>Edges 19490 | 0.001 | Performance Management | SNA, Performance, Innovation, Sustainability, KM |
| Orange | Nodes 5674<br>Edges 21075 | 0.001 | Machine Learning | Classification, Ensemble learning, Extreme learning machines, Genetic algorithms, Semi-supervised learning |
| Green | Nodes 4788<br>Edges 17749 | 0.002 | AI for System Identification | Neural networks, Adaptive control, Fuzzy logic, Nonlinear system, Uncertainty |

The keyword network is divided by the use and thematic function of keywords in HR. An example would be AI for system identification and reasoning control, which holds a significant position within the green cluster of AI keywords. The next steps include adaptive control and system identification, which are crucial uses of AI. Therefore, naming the clusters involves studying the impact of keywords on HR and the interaction of the biggest nodes [14, 101].

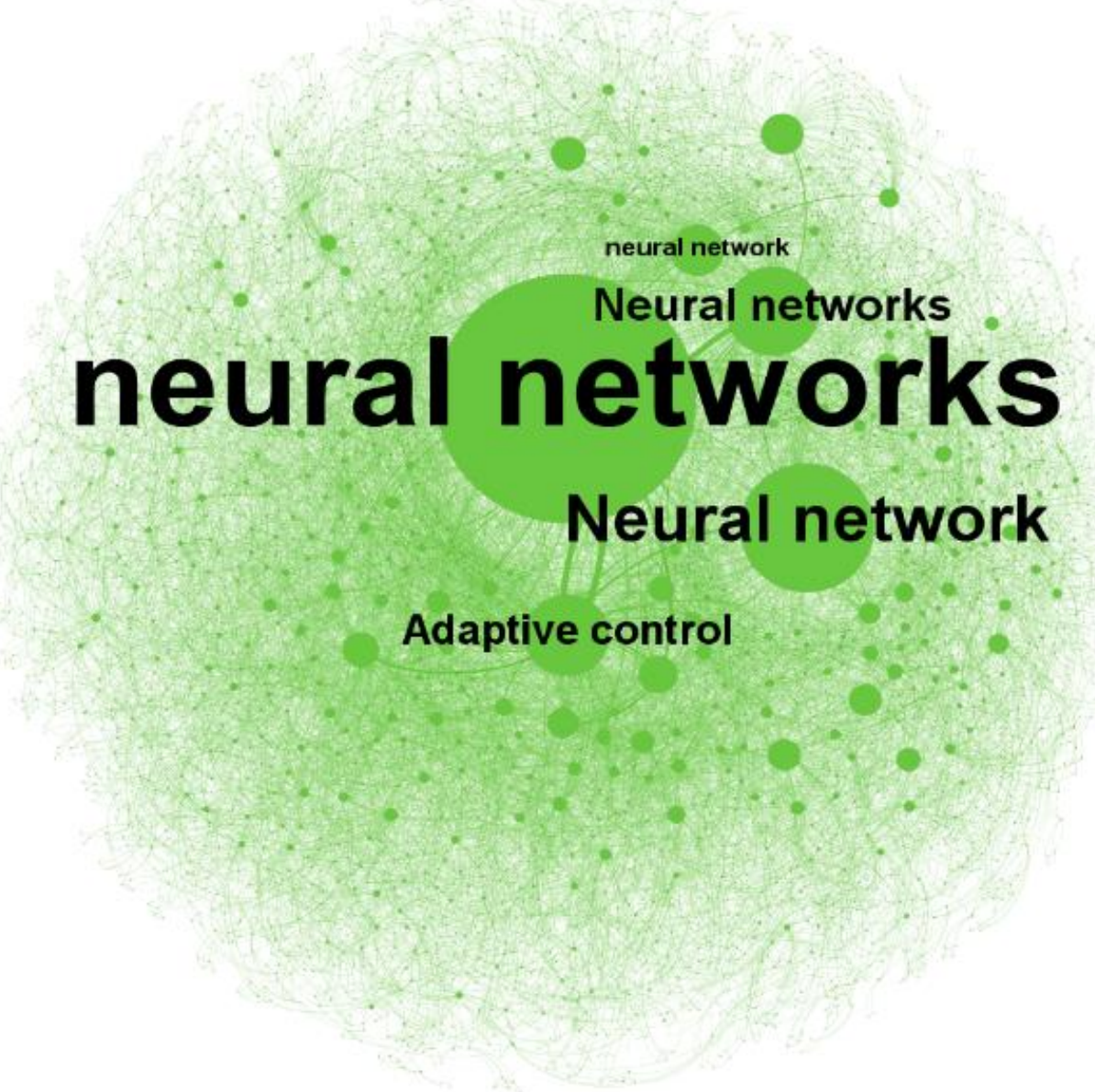

Figure 15 Green keywords community

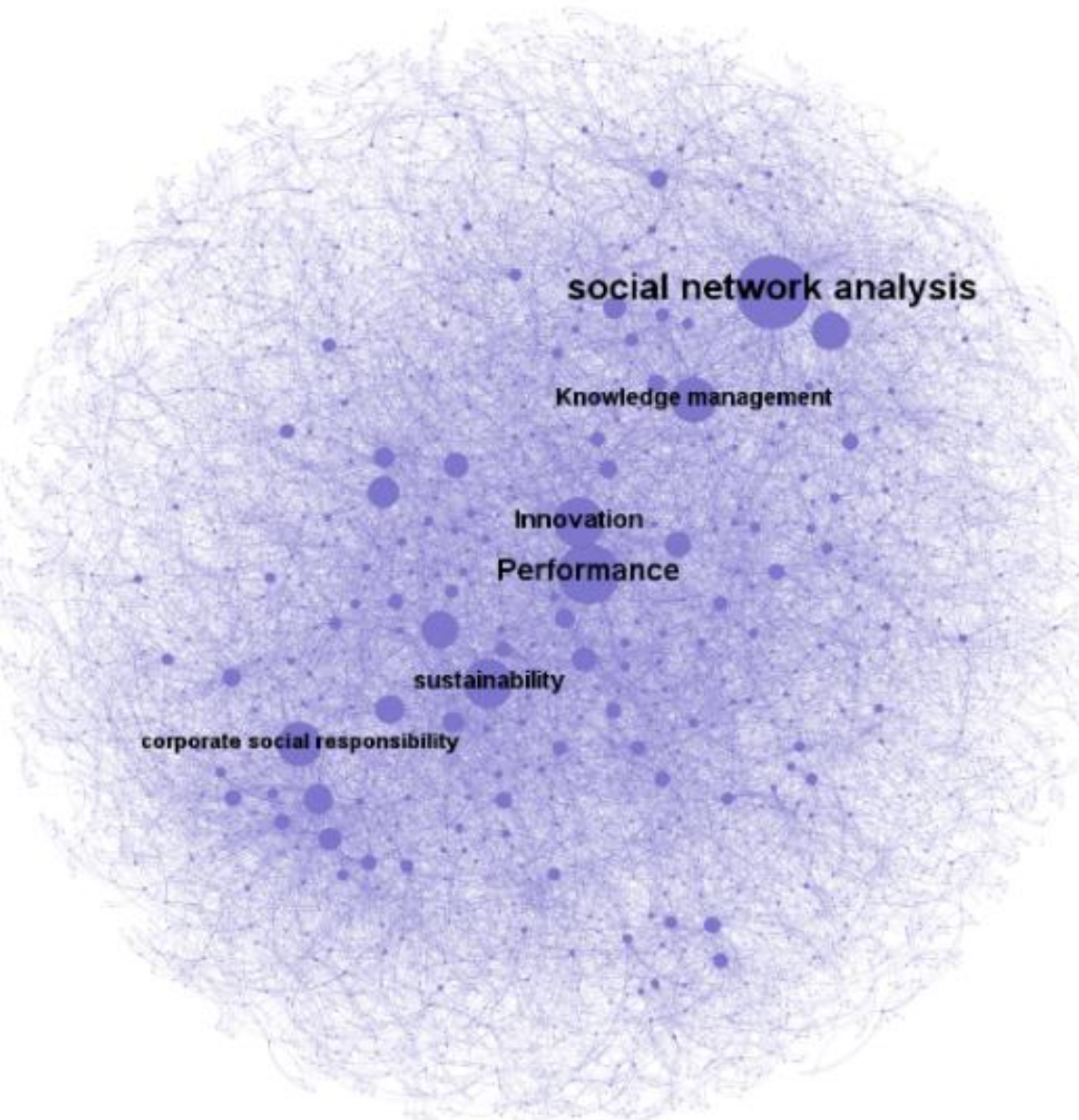

Figure 16 Purple keywords community

**AI for System Identification and Control**: The eco-friendly community, depicted in Figure 15 , constitutes a collective of researchers and experts who share a passion for exploring novel applications of AI in HR. The community's focus on neural networks, adaptive control, and system identification underscores the significance of these areas in implementing AI in HR management. By exploring the potential of these techniques, the community is seeking to uncover innovative solutions that can revolutionize HR processes. One of the primary areas of interest within this community is the development of neural networks, which can be used to model complex HR datasets and make predictions about employee behavior and performance. By leveraging adaptive control systems, HR professionals can adjust and refine their strategies in real-time, ensuring that HR processes are dynamic and responsive to changing organizational needs. System identification, another key area of focus, involves analyzing and understanding HR systems and processes to identify areas for improvement. By applying AI-powered analytics and machine learning algorithms to HR data, the community is working to develop more accurate and effective solutions for HR management.

The community's emphasis on stability and resilience also highlights the importance of creating HR systems that are robust and able to withstand the complexities and uncertainties of modern organizations. By developing AI-powered HR systems that are designed to be stable and resilient, HR professionals can ensure that HR processes are efficient, effective, and able to adapt to changing circumstances.

Overall, the eco-friendly community's research and development efforts aim to create a new generation of AI-powered HR management systems that are more efficient, effective, and responsive to the needs of modern organizations. By exploring the potential of neural networks, adaptive control, and system identification, the community is working to redefine the future of HR management and create a more sustainable, equitable, and prosperous work environment for all [102].

**HR Analytics and Performance Management**: The purple group Figure 16 has identified key areas of focus in HR, prioritizing performance, innovation, sustainability, knowledge management, and corporate social responsibility. Within these areas, HR Analytics and Performance Management play a crucial role. By leveraging HR analytics, organizations can make data-driven decisions, optimize HR processes, and drive business outcomes. Meanwhile, performance management is essential for setting goals, measuring progress, and ensuring accountability. By aligning these areas, HR professionals can create a more effective, efficient, and strategic HR function that supports the organization's overall objectives.

The purple group's focus on small and medium-sized enterprises (SMEs) is particularly noteworthy, as it highlights the importance of utilizing AI to enhance and simplify HR processes in companies of all sizes. By leveraging AI-powered HR solutions, SMEs can streamline tasks, reduce administrative burdens, and make more informed decisions. Moreover, AI can help SMEs to stay competitive in the marketplace, attract and retain top talent, and improve employee engagement. By focusing on these key areas, the purple group is poised to make a significant impact on the HR industry, driving innovation, and shaping the future of HR management for years to come[103].

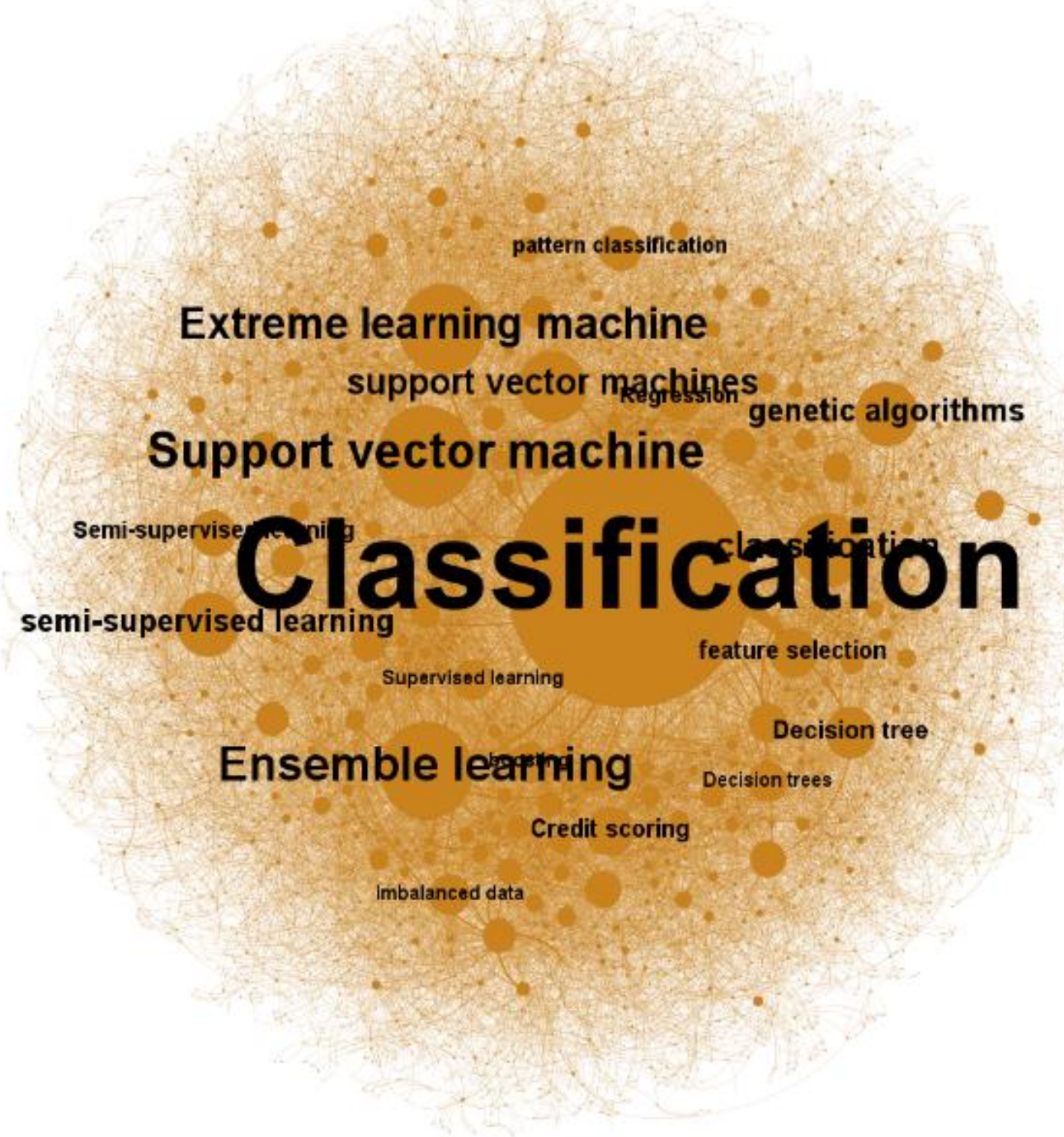

Figure 17 Orange keywords community

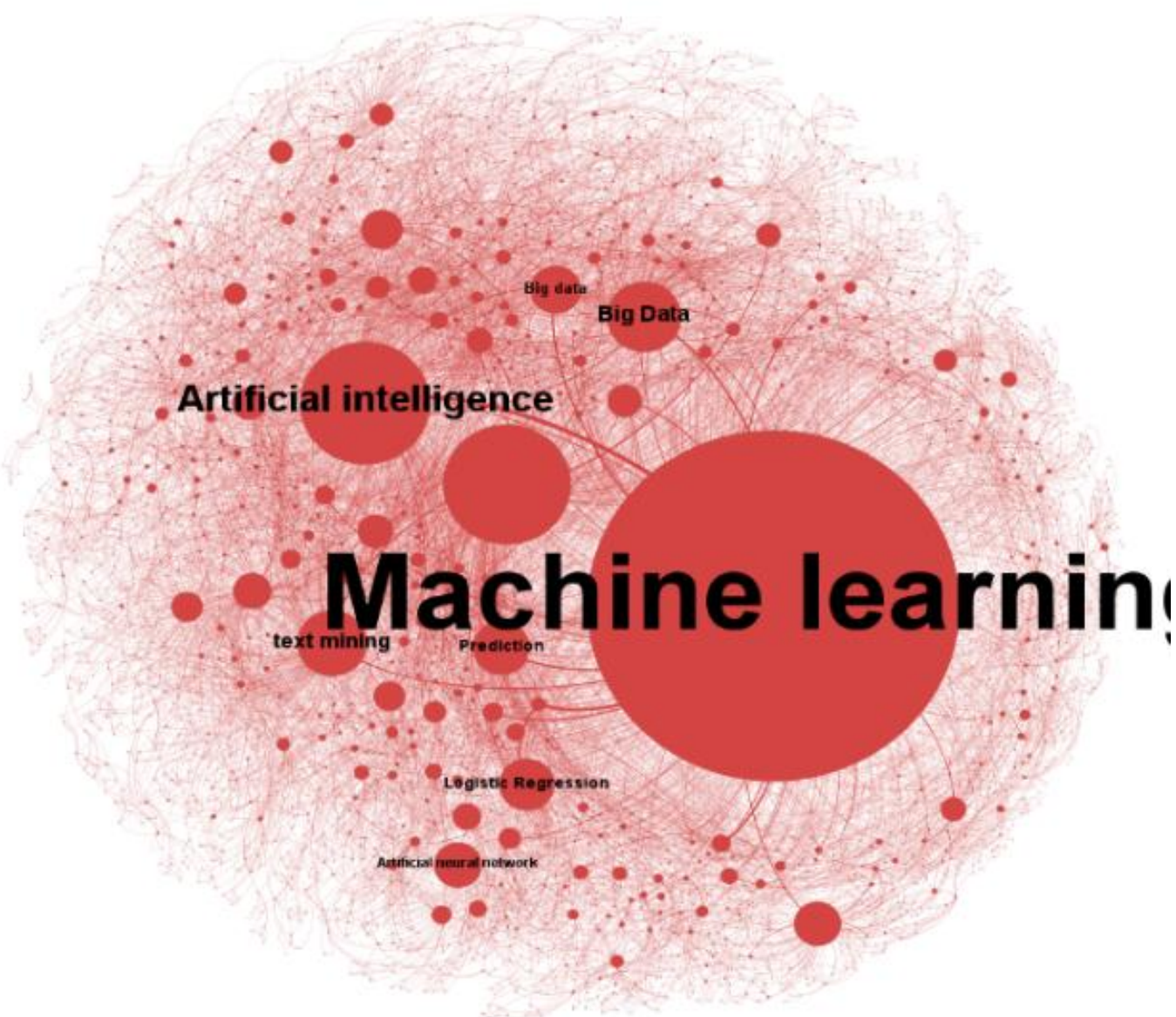

Figure 18 Red keywords community

**Machine Learning for Classification and Prediction**: The orange community focuses in Figure 17 is dedicated to exploring the applications of machine learning techniques in various domains, including classification, regression, and ensemble learning. These methods have far-reaching implications in HR management, enabling organizations to make informed decisions, identify trends, and predict outcomes. For instance, classification techniques can be used to identify high-potential staff, while regression methods can help predict employee turnover. Ensemble learning approaches, which combine multiple models, can enhance the accuracy and reliability of predictions.

The orange community's focus on skewed data and class imbalance is particularly noteworthy, as it highlights the challenges and opportunities that arise when dealing with imbalanced datasets. In HR management, skewed data can occur when there are more instances of one class (e.g., employees who have left the company) compared to another class (e.g., employees who are still with the company). By developing techniques to address skewed data and class imbalance, the orange community can create more effective machine learning models that accurately predict HR outcomes. The potential benefits of this research are vast, including improved staff retention, enhanced talent acquisition, and more informed HR decision-making[104].

**AI and HR in Decision-Making**: Figure 18 depicts a network of interconnected concepts related to AI, machine learning, and big data. Key elements like AI, Machine Learning, and Big Data are prominently featured. Each of these elements is crucial for harnessing the power of data and extracting insights that can inform business decisions. The presence of subjects like Human Resource Management (HRM), intellectual capital, and interpretability further highlights the group's focus on applying AI to support HR tasks.

The group's emphasis on HRM and intellectual capital suggests a keen interest in leveraging AI to streamline HR processes and unlock the full potential of intellectual capital within organizations. By harnessing the power of AI, HR professionals can gain strategic insights that inform talent management, employee engagement, and overall business performance. The incorporation of interpretability also indicates a desire to ensure that AI-driven insights are transparent, explainable, and actionable, enabling HR professionals to make informed decisions that drive business outcomes. Overall, Figure 18 provides a compelling glimpse into the group's ambition to harness the power of AI to revolutionize HR practices and drive business success. [17].

## V. DISCUSSION

This study examined the co-authorship network within AI applications in HRM research using social network analysis. The findings provide insights into the key players, collaboration patterns, and emerging themes in this rapidly evolving field. Here we discuss the results in relation to the four research questions posed:

***Q1) Who are the most influential authors in the HRM and AI co-authoring network?***

The analysis identified several highly influential authors based on centrality measures, with Pedrycz Witold, Chen C.L. Philip, and Wang Wei emerging as the top three most central authors. These researchers demonstrate strong collaborative ties and play pivotal roles in connecting different research communities within the AI-HRM domain. Their prominence suggests they are driving much of the innovative work bridging AI and HRM.
Interestingly, while these top authors come from diverse geographical backgrounds (Canada, China, and the USA respectively), there is a notable concentration of influential researchers from China and other Asian countries. This aligns with the rapid growth of AI research and implementation in these regions. The diversity of top authors' expertise - ranging from computational intelligence to data mining and strategic management - highlights the interdisciplinary nature of AI applications in HRM.

***Q2) Is the collaboration between countries in co-authorship driven by shared challenges, and what underpins scientific collaboration among universities and institutions in HRM and AI?***
The analysis revealed distinct collaboration communities among countries, suggesting that shared regional challenges and research priorities do influence co-authorship patterns. For instance, the "Global HR Applications" community centered around European countries indicates a common focus on integrating AI into HR practices within the EU regulatory framework. Similarly, the "HRM in the Middle East and Asia" community reflects shared interests in leveraging AI for HR in emerging economies.
Institutional collaboration appears to be driven by several factors:
- Geographic proximity, as seen in the clustering of Asian research centers
- Shared research interests, exemplified by the community focused on AI in higher education
- Access to resources and expertise, with leading AI research institutions forming collaborative networks

These collaboration patterns suggest that while global cooperation exists, there are still regional and institutional clusters that may benefit from increased cross-pollination of ideas.

***Q3) What underpins scientific collaboration among universities and institutions in HRM and AI?***
The analysis identified five main communities in the institutional co-authorship network, each with distinct characteristics:
- AI Research Centers Dominant in Asia: Focused on adapting Western HRM practices and integrating AI
- AI in Leading US Universities: Addressing labor shortages and virtual team management
- Leaders in AI Research in Europe: Emphasizing more informal, culturally-aligned HRM practices
- AI in Higher Education on a Global Scale: Exploring cultural variations in AI-driven HRM
- Newly Established AI Research Centers: Representing a mix of developing and progressing economies

These communities reflect different priorities and approaches to AI in HRM, influenced by factors such as cultural context, economic development stage, and specific organizational challenges. The diversity of these communities underscores the need for contextualized approaches to AI implementation in HRM across different global regions.

***Q4) Does the analysis of keyword networks in HRM and AI research reveal distinct and coherent patterns?***
The keyword network analysis revealed four distinct communities, each representing a coherent research theme:
1. AI for System Identification and Control: Focusing on neural networks and adaptive control in HR processes
2. HR Analytics and Performance Management: Emphasizing the use of AI for enhancing HR processes, particularly in SMEs
3. Machine Learning for Classification and Prediction: Exploring applications like identifying high-potential staff or predicting turnover
4. AI and HR in Decision-Making: Concentrating on the integration of AI, machine learning, and big data for HR decision support

These themes provide a clear roadmap of the current focus areas in AI-HRM research. They suggest that the field is progressing from basic implementation of AI tools to more sophisticated applications that can potentially transform HR decision-making and strategic planning.
The emergence of these distinct yet interrelated themes indicates that AI in HRM is a multifaceted field, with researchers exploring various aspects from technical implementation to strategic impact. This diversity of research focus areas suggests a rich and evolving landscape of AI applications in HRM, with ample opportunities for further innovation and cross-pollination of ideas across these thematic communities.

### *A. Managerial Insights and Implication*
The findings from this social network analysis of AI applications in HRM research offer valuable insights for managers and practitioners seeking to leverage AI technologies in their HR practices. These insights can guide implementation strategies and help organizations navigate the complex landscape of AI-driven HRM.

**1. Strategic Collaboration and Knowledge Sharing:**

The identification of key influential authors and institutions provides a roadmap for managers looking to stay at the forefront of AI-HRM innovations. Managers should:

- Actively seek collaborations or partnerships with leading institutions and researchers in the field.
- Encourage their HR teams to engage with academic research, perhaps through industry-academia partnerships or sponsored research programs.
- Implement knowledge-sharing mechanisms within their organizations to disseminate insights from cutting-edge research.

**2. Regional Adaptation of AI-HRM Practices:**

The analysis revealed distinct regional patterns in AI-HRM research, suggesting that one-size-fits-all approaches may not be optimal. Managers should:

- Tailor AI-HRM implementations to local contexts, considering cultural nuances, regulatory environments, and regional challenges.
- Look to region-specific research clusters for insights that may be most relevant to their operational context.
- Balance global best practices with localized approaches, especially in multinational corporations.

**3. Focus on Emerging Themes:**

The keyword analysis highlighted four main themes in AI-HRM research. Managers can use these to prioritize their AI initiatives:

**a) AI for System Identification and Control**

This theme focuses on leveraging AI technologies for systematic workforce management and organizational control mechanisms:

- o Workforce Planning Systems: Advanced AI algorithms analyze historical data, market trends, and organizational requirements to optimize resource allocation and workforce distribution. These systems can predict staffing needs across departments and projects with greater accuracy.
- o Adaptive Performance Management: Implementation of real-time monitoring and feedback systems that automatically adjust performance metrics based on changing business conditions. This includes dynamic goal-setting and performance evaluation frameworks that evolve with organizational needs.
- o Competency Gap Analysis: AI-powered systems that continuously assess employee skills against evolving job requirements, identifying training needs and development opportunities proactively. These systems can predict future skill requirements and suggest targeted learning interventions.

**b) HR Analytics and Performance Management**

This theme encompasses the integration of advanced analytics into core HR functions:

- o Data Infrastructure Development: Implementation of robust data collection and storage systems capable of handling diverse HR data types, from performance metrics to employee feedback. This includes establishing data governance frameworks and ensuring data quality.
- o Real-time Performance Analytics: Development of sophisticated dashboards that provide instant insights into employee performance, team dynamics, and organizational health metrics. These systems enable proactive intervention and continuous improvement.
- o Personalized Development Tracking: AI-driven platforms that create and monitor individualized employee development plans, tracking progress and automatically adjusting learning paths based on performance data and career aspirations.

**c) Machine Learning for Classification and Prediction**

This theme focuses on applying ML algorithms to enhance HR decision-making processes:

- o Advanced Talent Acquisition: Implementation of sophisticated ML models that analyze candidate profiles, job requirements, and organizational culture to predict job fit and success probability. These systems consider both technical skills and soft competencies.
- o Retention Risk Analytics: Development of predictive models that identify flight risks by analyzing patterns in employee behavior, engagement levels, and external market conditions. These systems enable proactive retention strategies.
- o High-potential Identification: ML algorithms that analyze multiple data points (performance history, leadership qualities, learning agility) to identify and track high-potential employees for succession planning.

**d) AI and HR Decision-Making**

This theme explores the integration of AI into strategic HR decision processes:

- o Strategic Planning Support: Implementation of AI systems that analyze workforce trends, market conditions, and organizational data to support strategic HR planning and policy development. These systems help identify long-term workforce needs and challenges.

- o Advanced Sentiment Analysis: Development of sophisticated NLP systems that analyze employee feedback, communication patterns, and social interactions to gauge organizational climate and employee satisfaction levels.
- o Objective Decision Frameworks: Implementation of AI-powered decision support systems that reduce bias in HR processes by standardizing evaluation criteria and providing data-driven recommendations for promotions, compensation, and career advancement.

**4. Interdisciplinary Approach:**
The diversity of backgrounds among top researchers suggests that effective AI-HRM implementation requires an interdisciplinary approach. Managers should:
- Form cross-functional teams that include HR professionals, data scientists, and IT specialists.
- Invest in upskilling HR staff in data literacy and basic AI concepts.
- Foster a culture of continuous learning and adaptation to keep pace with AI advancements.

**5. Bridging Academia and Industry:**
The strong presence of academic institutions in the network suggests a wealth of theoretical knowledge that may not yet be fully translated into practice. Managers should:
- Establish partnerships with universities for applied research in AI-HRM.
- Offer internships or collaborative projects to bridge the gap between academic research and practical implementation.
- Contribute to academic research by sharing real-world implementation challenges and successes.

the successful implementation of AI in HRM requires a strategic, multifaceted approach. By leveraging insights from the research landscape, fostering collaborations, and maintaining a balance between innovation and ethical considerations, managers can position their organizations to fully harness the potential of AI in HRM. The key lies in viewing AI not as a replacement for human judgment, but as a powerful tool to augment and enhance HR capabilities, ultimately leading to more data-driven, efficient, and employee-centric HR practices.

### *B. Limitations and Future Research*

Despite the valuable insights provided by this study, there are several limitations to consider. First, the analysis relies primarily on co-authorship networks and citation data from the Web of Science database, which may not fully capture the breadth of research activities, particularly from grey literature or non-English publications. Additionally, the study's focus on quantitative metrics such as centrality measures and TOPSIS may overlook qualitative aspects of collaboration dynamics, such as the depth of research partnerships or the nature of interdisciplinary work. Finally, the dataset, while extensive, may have limitations in terms of representing emerging trends in AI-HRM research, as newer works may not yet be fully captured in citation data.

Future research could address the limitations of this study by incorporating alternative sources of data, such as patents, conference proceedings, or industry reports, to provide a more comprehensive view of AI applications in HRM. Furthermore, qualitative studies exploring the nature of collaboration, including interviews with key authors and institutions, could provide deeper insights into the mechanisms of research partnerships. In addition, expanding the focus to include the practical applications of AI in HRM within organizations and its impact on workforce outcomes would enrich the understanding of AI's role in HRM. Longitudinal studies tracking the evolution of AI-HRM integration over time would also be valuable to capture the dynamic changes in the field and their implications for HR practices.

## VI. CONCLUSIONS

This study provides a comprehensive analysis of the research landscape in AI-HRM using social network analysis. By examining co-authorship networks, institutional collaborations, and keyword patterns, this research offers valuable insights into the structure and dynamics of this rapidly evolving field.
The analysis reveals a complex and interconnected research ecosystem, with several key findings:

- **Influential Authors:** The study identifies a core group of highly influential researchers, led by Pedrycz Witold, Chen C.L. Philip, and Wang Wei, who play pivotal roles in shaping the field. These authors, from diverse geographical and academic backgrounds, demonstrate the interdisciplinary nature of AI-HRM research and highlight the global reach of this emerging field.
- **Global Collaboration Patterns:** The research uncovers distinct collaboration communities among countries, reflecting shared regional challenges and research priorities. The identification of three main communities - Global HR Applications, HRM in the Middle East and Asia, and Global Integration of AI in HRM - illustrates the diverse approaches to AI-HRM across different global regions and emphasizes the importance of contextual factors in shaping research agendas.
- **Institutional Networks:** The analysis of institutional co-authorship reveals five distinct communities, each with unique characteristics and focus areas. This finding underscores the varied approaches to AI-HRM research across different academic and cultural contexts, from the adaptation of Western HRM practices in Asia to the exploration of AI in higher education globally.

- **Emerging Research Themes:** The keyword network analysis identifies four primary research themes: AI for System Identification and Control, HR Analytics and Performance Management, Machine Learning for Classification and Prediction, and AI and HR in Decision-Making. These themes provide a roadmap for future research and highlight the multifaceted nature of AI applications in HRM.

The study's findings have significant implications for both academia and industry:

- **For researchers**, this study provides a comprehensive overview of the current state of AI-HRM research, identifying key players, collaboration opportunities, and emerging themes. This can guide future research directions and foster more targeted collaborations across institutions and regions.
- **For practitioners**, the identified research themes offer insights into the most promising areas for AI application in HRM. This can help organizations prioritize their AI investments and implementation efforts, aligning their initiatives with cutting-edge research.
- **For policymakers**, This study offers valuable insights for policymakers by elucidating the impact of AI-driven human resource management (HRM) practices on organizational efficiency and workforce management. By mapping the emerging trends and collaboration patterns in AI-HRM research, policymakers can gain a clearer understanding of the potential benefits and challenges associated with integrating AI in HRM practices. These insights can guide the development of regulatory frameworks and public policies that support the responsible and effective adoption of AI technologies, ensuring that AI integration in HRM aligns with national workforce development goals, improves productivity, and mitigates potential risks such as job displacement or skill gaps.

In conclusion, this study offers a thorough and insightful mapping of the AI-HRM research landscape, shedding light on the prevailing trends, influential contributors, and emerging directions in the field. As artificial intelligence progressively reshapes human resource management practices, this research serves as a crucial resource for scholars, practitioners, and policymakers. For policymakers, the findings guide the formulation of strategies that foster international collaboration by pinpointing key research hubs and partnerships. Furthermore, the study identifies critical areas where AI implementation in HRM necessitates careful consideration of ethical and regulatory issues. It underscores the importance of ensuring responsible deployment of AI technologies, promoting innovation while addressing challenges related to bias, fairness, and privacy. Ultimately, this study lays the groundwork for navigating the dynamic intersection of AI and HRM, encouraging global collaboration and responsible AI integration in HR practices. Future research should expand on these findings, delving into the qualitative aspects of AI-HRM implementation, exploring ethical dilemmas, and evaluating the long-term implications of AI on workforce dynamics and organizational performance. Additionally, longitudinal studies are essential to track the evolution of this field, providing a deeper understanding of how AI-HRM research and practice continue to shape the future of work.

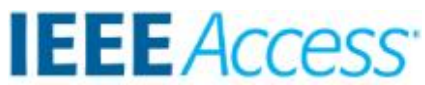